\documentclass[11pt]{article}
\usepackage[autonum,bbold,colorhypersetup]{shortex}
\usepackage{url}
\usepackage{amsfonts}
\usepackage{amsmath}
\usepackage{amssymb}
\usepackage{amsthm}
\usepackage{array}
\usepackage{arydshln}
\usepackage{bm}
\usepackage{booktabs}
\usepackage[dvipsnames]{xcolor}
\usepackage{easybmat}
\usepackage{graphicx}
\usepackage{times}
\usepackage[round]{natbib}
\usepackage[margin=1in]{geometry}
\usepackage{setspace}
\usepackage[titletoc,title]{appendix}

\usepackage{color}

\def\blue#1{\textcolor{blue}{#1}}

\makeatletter
\def\blfootnote{\gdef\@thefnmark{}\@footnotetext}
\makeatother

\title{Margin-closed regime-switching multivariate time series models}
\author{\normalsize
Lin Zhang$^{1\ast}$\qquad
Harry Joe$^{1}$\qquad
Natalia Nolde$^{1}$\\
{\footnotesize  $^1$\textit{Department of Statistics, University of British Columbia,
Vancouver, BC Canada V6T 1Z4}}
}
\date{}

\begin{document}
\blfootnote{$^{\ast}$Corresponding author. Email address: lin.zhang@stat.ubc.ca}
\maketitle

\begin{abstract}

A regime-switching multivariate time series model which is closed under margins is built. The model imposes a restriction on
all lower-dimensional sub-processes to follow a regime-switching
process sharing the same latent regime sequence and having the same Markov order as the original process. 
The margin-closed regime-switching model is constructed by considering
the multivariate margin-closed Gaussian VAR($k$) dependence as
a copula within each regime, and builds dependence between observations in
different regimes by requiring the first observation in the new regime
to depend on the last observation in the previous regime.
The property of closure
under margins allows inference on the latent regimes based on lower-dimensional selected sub-processes and estimation of univariate parameters
from univariate sub-processes, and enables the use of multi-stage estimation procedure for the model.
The parsimonious dependence structure of the model also avoids a large number of parameters under the regime-switching setting. The proposed model is applied to a macroeconomic data set to infer the latent business cycle and compared with the relevant benchmark.

%

\medskip


\end{abstract}

\section{Introduction}
\label{sec:introduction}


In this paper we consider the setting in which data are in the form of a multivariate discrete-time time series with observations on $d$ variables. The time series is non-stationary due to latent states, and the state of each regime remains constant over prolonged periods of time. One example for such a setting is the series of multiple macroeconomic indicators whose behavior is influenced by the business cycle. The latent states in this case correspond to economic expansion and recession.  

To model this type of time series data and infer the underlying latent regime sequence, we propose a regime-switching model that aims to balance flexibility, interpretability and parsimony in view of high computational costs associated with model fitting. In particular, we allow for flexible modeling of the marginal univariate distributions within each regime. A simple serial dependence construction is suggested for transitions between regimes. Within each regime, the dependence structure of the Gaussian vector autoregressive (VAR) time series model is imposed. But an important contribution to the existing literature includes the restriction of the entire process to be closed under margins. In the context of a stationary VAR model, closure under margins is considered in \citet{Zhang2023}. This restriction for the regime-switching model reduces the number of parameters, allows inference on the latent regimes based on lower-dimensional selected sub-processes and estimation of univariate parameters from univariate sub-processes.


There is related literature on regime-switching models with conditional serial dependence within states; see \cite{Cheng2016}, \cite{Sola1994}, and \cite{Hamilton1990}. These papers introduce VAR processes with time-varying coefficient matrices. These models do not handle non-Gaussian stationary distributions within regimes. To reduce the number of parameters in these models, \cite{Monbet2017} propose to adopt penalization of the likelihood to reduce the number of parameters by shrinking estimates for some of them to zero. These papers do not consider the behavior of the model for marginal sub-processes.

The remainder of the paper is organized as follows.
\cref{sec:mc_VAR} reviews background on the margin-closed VAR models
in preparation for
\cref{sec:mc_rs_model} which provides details of the proposed
margin-closed regime-switching multivariate time series model and its parameterization.
\cref{sec:fitting_models} discusses estimation of the margin-closed regime-switching model and inference for the latent regime sequence.
\cref{sec:simulation} presents a simulation study to show the effect
of different location shifts and  different dependence structures on
the model fitting and latent regime inference.
\cref{sec:empirical} illustrates the proposed methodology on a macroeconomic data set.
\cref{sec:discussion} contains final remarks. The Appendix includes the proof of the closure under margins result in \cref{sec:mc_rs_model}, as well as several other supplementary derivations.

\section{Margin-closed Gaussian VAR model}
\label{sec:mc_VAR}

The main focus of this paper is the proposal of a
regime-switching multivariate time series model that is based on a
Gaussian stationary VAR time series model within each regime.
This section provides the background on the margin-closed Gaussian VAR model and reviews how it can be parameterized; for details, see \cite{Zhang2023}.

Let the dimension of the multivariate time series be $d$.
For the observed $d$-variate time series, the stationary distribution in each regime need not be Gaussian. Within every regime, each of the $d$ continuous random variables 
is transformed via the univariate probability integral transform
to standard Gaussian. Our assumption is that the joint distribution of all consecutive sequences
of the $d$-dimensional random vectors $\cbra{\bZ_t}_{t\in\mathbb{N}}$ of these transformed variables is multivariate Gaussian.
This approach is used in \cite{biller2003modeling} and
\cite{biller2009copula} to define time series models with non-Gaussian stationary margins.

Let $\cbra{\bZ_t}_{t\in\mathbb{N}}$ with $\bZ_t =\rbra{Z_{1,t},\dots,Z_{d,t}}^\top$ denote a $d$-variate standardized Gaussian time series; i.e., each component $Z_{i,t}$ ($i=1,\ldots,d$) has zero mean and unit variance.
Then the Gaussian VAR($k$) time series model has the following stochastic representation:
\begin{equation}\label{eq:VAR_model}
  \bZ_t = \bPhi_1\bZ_{t-1} + \cdots + \bPhi_k\bZ_{t-k} + \bepsilon_t, \quad \bepsilon_t\overset{\mathrm{i.i.d.}}{\sim} \distNorm_d\rbra{\bm{0}, \bSigma_{\bepsilon}},
\end{equation}
where $\bPhi_1,\ldots, \bPhi_k$ denote $d\times d$ coefficient matrices. 
Let $I_d$ denote the identify matrix of dimension~$d$.
When the coefficient matrices  $\{\bPhi_j: j=1,\dots,k\}$ in \cref{eq:VAR_model} satisfy the stationarity condition that $\mathrm{det}\rbra{I_d-\bPhi_1z-\cdots-\bPhi_k z^k}\neq 0\ \mbox{for}\ |z|<1$, the VAR($k$) process is characterized by the stationary joint distribution of $(k+1)$ consecutive observations, i.e., the joint distribution of $\rbra{\bZ_t^\top,\dots,\bZ_{t-k}^\top}^\top$.
The VAR($k$) model for $\cbra{\bZ_t}_{t\in\ints}$ can be specified by the block Toeplitz correlation matrix $R_{\bZ_{t:(t-k)}}=\corr\rbra{\rbra{\bZ_t^\top,\dots,\bZ_{t-k}^\top}^\top}$.
The subscript notation $A_{t_a:t_b}$ used here and subsequently is an abbreviation for
$(A_{t_a},A_{t_a-1},\ldots,A_{t_b})$ if $t_a>t_b$.

A VAR($k$) process is closed under margins with respect to a partition if
and only if every sub-process in the partition is also a multivariate
VAR($k$) process of lower dimension or a univariate autoregressive AR($k$) process.
For example, a trivariate VAR($k$) process $\cbra{\rbra{Z_{1,t},Z_{2,t},Z_{3,t}}}_{t\in\ints}$ is closed under margins with respect to partition $\cbra{\{1,2\},\{3\}}$ if and only if sub-processes $\cbra{\rbra{Z_{1,t},Z_{2,t}}}_{t\in\ints}$ and $\cbra{\rbra{Z_{3,t}}}_{t\in\ints}$ follow a bivariate VAR($k$) model and an AR($k$) model, respectively.
A special case of closure under margins is when the number of sub-processes in the partition is exactly the dimension of the original VAR($k$) process; that is, all sub-processes are univariate.

For VAR models, there are several advantages of the subclass with the closure under margins property.
First, the marginal models of any sub-process of a multivariate time series can be obtained by extracting the relevant parameters from the correlation matrix parametrization of the VAR model that the original multivariate time series follows.
For fitting VAR($k$) models, if all univariate components of the multivariate time series follow AR($k$) models, the AR($k$) models of all univariate components can be fitted first, followed by estimation of the cross-correlation parameters that are contemporaneous or lagged.
Intuitively, a VAR($k$) process is closed under margins if all coefficient matrices of the VAR($k$) model are diagonal.
\cite{Zhang2023} derived a sufficient condition under which a VAR($k$) model is closed under margins and showed that the coefficient matrices can be non-diagonal.
Furthermore, under a certain constraint, the serial correlations of all univariate components and the contemporaneous correlations between the univariate components can characterize a VAR($k$) model that is margin-closed with respect to any partition of the VAR($k$) process, and
coefficient matrices in \cref{eq:VAR_model} can be non-diagonal.

More specifically, if all univariate components $\cbra{Z_{i,t}}_{t\in\ints}$ for
$i=1,\dots,d$ are AR($k$),
the Toeplitz correlation matrix
$R_{Z_{i,t:(t-k)}}=\corr\rbra{(Z_{i,t},\dots,Z_{i,t-k})^\top}$ characterizes
the serial correlations of $\cbra{Z_{i,t}}_{t\in\ints}$.
Then, the correlation matrices $R_{Z_{i,t:(t-k)}}$ for
$i=1,\dots,d$, and the contemporaneous correlation matrix
$R_{\bZ}=(\corr\rbra{Z_{i,t},Z_{j,t}})_{1\leq i,j \leq d}$ can parameterize a
correlation matrix $R_{\bZ_{t:(t-k)}}$ such that
the VAR($k$) process $\cbra{\bZ_t}_{t\in\ints}$ is margin-closed with respect to
all partitions and its coefficient matrices can all be non-diagonal, under the
constraint that $R_{\bZ_{t:(t-k)}}$ is positive definite.
The approach to derive $R_{\bZ_{t:(t-k)}}$ from $R_{Z_{i,t:(t-k)}}$ for
$i=1,\dots,d$ and $R_{\bZ}$ is given in \cref{ap: mcvar}.
For details of the derivation, see \cite{Zhang2023}.


\section{Margin-closed regime-switching time series model}
\label{sec:mc_rs_model}

This section introduces a new regime-switching time series model with closure under margins.
After each variable in each regime has been transformed via the probability integral
transform to have the standard Gaussian distribution, 
margin-closed stationary Gaussian VAR time series models are assumed for each regime. 
In order to allow for dependence between observations before and after a
regime change, a parsimonious dependence assumption is used for this
transition. It is a minor generalization of a regime-switching model that
assumes the sequences of observations in different regimes are
mutually independent.
\cref{sec:model_formulation} specifies the model, and
\cref{sec:model_params} details on model parameterization and examples.

\subsection{Model formulation} \label{sec:model_formulation}

Consider a $d$-variate time series $\cbra{\bX_t}_{t\in\ints}$, where $\bX_t =
\rbra{X_{1,t},\dots,X_{d,t}}^\top$, and let
$\bx_t=(x_{1t},\dots,x_{dt})^\top$ and $v_t$ be the realization of $\bX_t$ and $V_t$ for $t\in\{1,\dots,T\}$.
Let $G$ be the total number of distinct latent regimes with $G\ll T$.
Let $V_t$ be the random variable specifying the regime at time $t$, and $v_t$ be its realization.
For a hidden Markov model with latent regimes or discrete states, conditional independence of the past beyond the previous time point is assumed, i.e., for any $t>1$,
\begin{equation}\label{eq:v_hmm}
  \sbra{V_{t}|V_{t-1}=v_{t-1},V_{t-2}=v_{t-1},\dots,V_1=v_1} \overset{\mathrm{d}}{=} \sbra{V_t|V_{t-1}=v_{t-1}}.
\end{equation}
Let $\bm{p}_V$ denote the $G$-dimensional vector with the initial probability
mass function of $V_1$, where the $g$-th element of $\bm{p}_V$ is equal to $\Pr(V_1=g)$ for $g\in\cbra{1,\dots,G}$.
Let $M_V$ be the $G\times G$ transition matrix of conditional probabilities.
Let $\btheta_V=\{\bm{p}_V, M_V\}$ denote the parameters of the hidden Markov model.
The process $\cbra{\bX_t}_{t\in\ints}$ is assumed to be strictly stationary within each regime.
Specifically, if the regime is $g\in\{1,\ldots,G\}$ for
$t_1<t<t_2$, then $\{\bX_t: t_1\le t<t_2, V_{t_1}=\cdots=V_{t_2-1}=g\}$ is stationary with
multivariate joint distribution $F_{\bX,g}$ having univariate margins $F_{i,g}$ for
the $i$-th variable, $i\in\{1,\ldots,d\}$.
Let $\bY_t=(Y_{1,t},\ldots,Y_{d,t})^\top$ with $Y_{i,t}$ obtained after probability integral transforms:
\begin{equation}
  Y_{i,t}=\Phi^{-1}\rbra{F_{i,g}\rbra{X_{i,t}}}
  =: a_{i,g}\rbra{X_{i,t}}, \quad
 1\leq i\leq d, \quad \mbox{for each}\ t\ \mbox{such that}\ V_t=g,
  \label{eq:PIT}
\end{equation}
where $\Phi$ is the standard Gaussian cumulative distribution function (CDF).
$\cbra{\bY_{t}}_{t\in\ints}$ can be regarded as a marginally transformed multivariate time series with $Y_{i,t} \sim \distNorm(0,1)$.

Next, we give a definition of our regime-switching model.

\begin{definition} \label{def:model}
(Regime-switching multivariate Gaussian time series model).
$\cbra{\bY_t, V_t}_{t\in\ints}$ satisfies the following conditions:

\medskip

\noindent\textbf{Condition 1.}
$\cbra{V_t}_{t\in\ints}$ is a Markov process, i.e., it satisfies~\cref{eq:v_hmm}.
$\cbra{\bY_t}_{t\in\ints}$ is a $d$-dimensional discrete-time Gaussian process
with $Y_{i,t}\sim \distNorm(0,1)$ for all $1\leq i\leq d$ and $t\in\ints$.

\medskip

\noindent\textbf{Condition 2.}
If $V_{t_1}=\cdots=V_{t_2-1}=g$
for $g\in\cbra{1,\dots,G}$, then
$\cbra{Y_{i,t}}_{t_1\leq t <t_2}$ follows an AR model with Markov order $k_{i,g}$ for $1\leq i\leq d$, and $\cbra{\bY_t}_{t_1\leq t <t_2}$ follows a margin-closed $d$-dimensional VAR($k_g$) model with Markov order $k_g=\max_{1\leq i \leq d}k_{i,g}$.

\medskip

\noindent\textbf{Condition 3.}
When regime transitions from $g$ to $g'$ at time $t=t^*$, i.e., $g=V_{t^*-1}\neq V_{t^*}=g'$, the following three assumptions are satisfied:
  \begin{enumerate}
  \item[(a)] The covariance matrix of last observation in one regime with the
  first observation in the next regime is:
  \begin{equation}
    \cov\rbra{\bY_{t^*}, \bY_{t^*-1}} = PR_{\bY,g},
    \quad V_{t^*}\ne V_{t^*-1}=g, \label{eq:condition1}
  \end{equation}
where $R_{\bY, g}=\corr\rbra{\bY_t|V_t=g}$,
$P=\diag\rbra{\rho_1, \dots, \rho_d}$ for $\rho_i\in(-1,1)$ for $1\leq i\leq d$.


  \item[(b)] The first observation in each regime is conditionally independent of all but the last observation in the previous regime given the last observation in the previous regime:
  \begin{equation}
   \sbra{\bY_{t^*}\bot \bY_{t^*-\ell-1}}\big|\bY_{t^*-1},\ \forall\ \ell\geq 1, \quad  V_{t^*}\ne V_{t^*-1}, \label{eq:condition2}
  \end{equation}

  \item[(c)] Observations in different regimes are independent except at the regime change point:
    \begin{equation}
   \bY_{t^*+\tilde{\ell}} \bot \bY_{t^*-\ell},\ \forall\ \ell, \tilde{\ell}\geq 1, \quad V_{t^*}\ne V_{t^*-1}. \label{eq:condition3}
  \end{equation}
  \end{enumerate}
\medskip
\end{definition}

Condition 3 specifies a parsimonious dependence assumption with
only $d$ extra parameters to handle transitions between regimes.
If $P$ is such that $\rho_1=\cdots=\rho_d=0$,
under the constraint of closure under margins,  the model reduces to
the simplest case of complete independence between observations in different regime periods.
Assumption (a) in \cref{eq:condition1} means that at a change point from time $t^*-1$ to $t^*$, the $d$ extra parameters $\rho_1,\dots,\rho_d$ such that
$\cov\rbra{Y_{i,t^*}, Y_{j,t^*-1}\big|V_{t^*}\neq V_{t^*-1}} = \rho_j\cov\rbra{Y_{i,t},Y_{j,t}| V_t=V_{t^*-1}}$.
For simplicity, these parameters do not depend on the actual states in the regime change, i.e., they are the same for regime $g$ to regime $g'$, or regime $g'$ to regime $g$.
For observations before time $t^*-1$, Assumption (b) in \cref{eq:condition2} indicates conditional independence between values of the process before time $t^*-1$ and at time $t^*$.
Assumption (c) in \cref{eq:condition3} further assumes that all observations after time $t^*$ are independent of observations before time $t^*-1$.
The three assumptions under Condition~3 above simplify the correlation structure at the time of regime switching by requiring that the first observation in the new regime only depends on the last observation in the previous regime, and other observations in the new regime are independent of the observations before regime switching.

In \cref{def:model}, note that (i) a univariate AR($k'$) time series
is AR($k$) with $k'<k$ if in the linear representation, the coefficients
for lags $k'+1$ to $k$ are 0 and the partial autocorrelations of
lags $k'+1$ to $k$ are 0; and (ii) a VAR($k'$) time series is
VAR($k$) with $k'<k$ if in the linear representation \cref{eq:VAR_model}, the coefficient
matrices for lags $k'+1$ to $k$ are zero matrices.
The following proposition shows that the process $\cbra{\bY_t}_{t\in\ints}$ specified in \cref{def:model}, although non-stationary  due to the changes of the regimes, is in fact still Markov of order $k$ and is closed under margins. 
\begin{proposition} \label{pro:markov_order}
If $\cbra{\bY_t,V_t}_{t\in\ints}$ satisfies \cref{def:model},
then $\cbra{\bY_t}_{t\in\ints}$ is Markov of order $k=\max_{1\leq g\leq G} k_g+1$ and is closed under margins, i.e., $\cbra{Y_{i,t}}_{t\in\ints}$ is Markov of order $k$ for any $1\leq i\leq d$,
and $\cbra{\bY_{I,t}}_{t\in\ints}$ is Markov of order $k$ where
$\bY_{I,t}=(Y_{i_1,t},\ldots,Y_{i_m,t})^\top$ with $I=\cbra{i_1,\ldots,i_m}$
being a subset of $\{1,\ldots,d\}$ with cardinality at least 2.
\end{proposition}
The proof is given in \cref{ap: proof}. 


Since $\cbra{\bY_t,V_t}_{t\in\ints}$ is a discrete-time Gaussian process,
there is a linear stochastic representation for $Y_t$ as a function
of $Y_{t-1},\ldots,Y_{t-k}$ that depends on the states $V_t,\ldots,V_{t-k}$.
\cref{example:mchmm} below presents a concrete case to illustrate the parameters in
a model satisfying \cref{def:model} and to show some resulting stochastic representations.

\subsection{Model parameterization} \label{sec:model_params}

In this section, details are provided for a parameterization of the regime-switching model in Definition~\ref{def:model}, and an example is given to illustrate calculation of correlation matrices and derivation of  stochastic representations for this time series model.

Consider the case that the process stays in the same regime from time~$t_1$ until $t_2-1$; i.e., $V_{t_1}=\cdots=V_{t_2-1}=g$ for some $g\in\cbra{1,\dots,G}$.
Since $\cbra{\bY_{t}}_{t_1\leq t <t_2}$ is a VAR($k_g$) process with $k_g<k$,
let
 \[R_g = \corr\rbra{\bigl(\bY_t^\top, \dots, \bY_{t-k}^\top\bigr)^\top
  \bigm| V_t=\cdots=V_{t-k}=g}\]
denote the $(k+1)d$-dimensional block Toeplitz correlation matrix for regime $g$.
Let the $(k+1)$-dimensional Toeplitz correlation matrix of the univariate components in regime $g$ be
  \[R_{i,g}=\corr\rbra{\bigl(Y_{i,t},\dots,Y_{i,t-k}\bigr)^\top|V_t=\cdots=V_{t-k}=g}\ \mbox{for}\ 1\leq i\leq d,\]
and let $R_{\bY,g}=\corr\rbra{\bY_t|V_t=g}$ be
the $d$-dimensional contemporaneous cross-sectional correlation matrix.
Note that $R_{i,g}$ can be extracted from $R_{g}$ via the rows/columns indexed by $(i,i+d,\ldots,i+kd)$.
With closure under margins,
$\cbra{Y_{i,t}: V_t=g}$ is an AR($k_{i,g}$) process, and
$R_{i,g}$ can be parameterized by the partial autocorrelations $\alpha_{i,\ell,g}\in (-1,1)$ of
lag $\ell\in\{1,\ldots,k\}$,
where
 \[\alpha_{i,\ell,g} = \corr\rbra{Y_{i,t}, Y_{i,t-\ell}|Y_{i,t-1},\dots,Y_{i,t-\ell+1}}\ \mbox{for}\ 1<\ell\leq k, \quad V_t=\cdots=V_{t-\ell}=g.\]
Note that $\alpha_{i,\ell,g} = 0$ for $k_{i,g}<\ell\leq k$. The entries of matrix $R_{\bY,g}$ are denoted by $\rho_{i,j,g}\in(-1,1)$ in the $i$-th row and $j$-th column for $1\leq i<j\leq d$, and the matrix is constrained to be positive definite.
Then, for a margin-closed VAR process in regime $g$, matrix $R_{g}$ can be parameterized by
$$ \rbra{\alpha_{i,1,g},\dots, \alpha_{i,k_{i,g},g} }\in (-1,1)^{k_{i,g}+1}\ \mbox{and}\ \rho_{i,j,g} \in (-1,1), \ \mbox{for }\ 1\le i<j \leq d,
  $$
with the constraint that $R_g$ is positive definite.

For dependence between consecutive observations at the times of regime switches, the $d$ extra parameters $\rho_1,\dots,\rho_d$ are adopted; see ~\cref{eq:condition1}. 

Since \cref{pro:markov_order} shows that $\cbra{\bY_t}_{t\in\ints}$ is Markov of order $k$, the likelihood of a realized time series $\cbra{\by_t: t=1,\ldots,T}$
is a product of conditional densities
$f_{\bY_t|\bY_{t-1},\ldots,\bY_{t-k}}(\by_t|\by_{t-1},\ldots,\by_{t-k})$
and only the joint distributions of $k+1$ consecutive values of $\cbra{\bY_t}_{t\in\ints}$ are needed.
Next we provide an expression for the correlation matrix of $\rbra{\bY_{t}^\top, \dots, \bY_{t-k}^\top}^\top$  that depends on the latent regimes
$V_t,\ldots,V_{t-k}$.
Let $R_{\bY_{t:(t-k)}|V_{t:(t-k)}=v_{t:(t-k)}}$ denote the
correlation matrix of $\rbra{\bY_{t}^\top, \dots, \bY_{t-k}^\top}^\top$ given $V_t=v_t,\dots, V_{t-k}=v_{t-k}$.
Let $\bSigma_{h_1,h_2,g}$ denote the
sub-matrix of first $h_1\times d$ rows and $h_2\times d$ columns of
$R_g$ for $1\leq h_1, h_2 \leq k+1$, i.e., $\bSigma_{h_1,h_2,g}$ is the
correlation matrix between $\rbra{\bY_{t}^\top, \dots,
\bY_{t-h_1}^\top}^\top$ and $\rbra{\bY_{t}^\top, \dots,
\bY_{t-h_2}^\top}^\top$ given $V_t=\cdots=V_{t-\max(h_1, h_2)}=g$.
Let $S\in\{0,\dots,k\}$ denote the number of regime switches from time points $t-k$ to $t$ with the $s$-th switch occurring at time $t_s$.
For $S=0$, regime does not switch in the time period $t-k$ to $t$ and $R_{\bY_{t:(t-k)}}=R_g$ for regime $g$.
For $S>0$, the time interval between $t-k$ to $t$ can be partitioned into $S+1$ periods such that the latent regime within a specific period is constant.
More specifically, the $S+1$ periods are discretized by time points $t_1,\dots,t_S$ with:
\begin{equation}
  \begin{gathered}
    v_{t_S}=\cdots=v_t = g_{S+1} ,\\
    v_{t_{S-1}}=\cdots=v_{t_S-1} = g_S \neq v_{t_S},\\
    \vdots\\
    v_{t-k}=\cdots= v_{t_1-1} = g_1 \neq v_{t_1}.
  \end{gathered}
\end{equation}
for some $t-k < t_1< t_2 < \cdots < t_S \leq t$, where $g_s$ is the regime of the $s$-th period.
Let $t_0 = t-k$, $t_{S+1}=t+1$, and define $e_s=t_s-t_{s-1}$ for $1\leq s\leq S+1$.
Let $\bOmega_{s}$ be the $e_{s+1}d\times e_sd$ covariance matrix between $\rbra{\bY_{t_s}^\top,\dots,\bY_{t_{s+1}-1}^\top}^\top$ and $\rbra{\bY_{t_{s-1}}^\top,\dots,\bY_{t_{s}-1}^\top}^\top$ for $1\leq s\leq S$. From \cref{eq:condition3}, $\cov(\bY_{t_s-\ell},\bY_{t_s+\tilde{\ell}})=\bm{0}$
for $\ell,\tilde{\ell}\ge 1$.
From \cref{eq:condition1}, $\cov(\bY_{t_s},\bY_{t_s-1})=PR_{\bY,g_s}
=P\cov(\bY_{t_s-1},\bY_{t_s-1})$.
From \cref{eq:condition2}, $\bY_{t_s}\perp\bY_{t_s-1-\ell}\mid \bY_{t_s-1}$
for $1\le\ell<t_s-t_{s-1}$ implies that
$$\bm{0}=\cov(\bY_{t_s},\bY_{t_s-1-\ell}) - \cov(\bY_{t_s},\bY_{t_s-1})[\cov(\bY_{t_s-1},\bY_{t_s-1})]^{-1}
\cov(\bY_{t_s-1},\bY_{t_s-1-\ell})$$
so that
\begin{multline}
  \cov(\bY_{t_s},\bY_{t_s-1-\ell}) = P\cov(\bY_{t_s-1},\bY_{t_s-1}) [\cov(\bY_{t_s-1},\bY_{t_s-1})]^{-1}
\cov(\bY_{t_s-1},\bY_{t_s-1-\ell}) \\= P\cov(\bY_{t_s-1},\bY_{t_s-1-\ell}).
\end{multline}
Combining the $e_s=t_s-t_{s-1}$ terms with $P$ leads to $\bOmega_{s} =
\rbra{\bm{0},\dots,\bm{0},(P\bSigma_{1,e_{s},g_s})^\top}^\top$.
Then, the $(k+1)d\times (k+1)d$ correlation matrix $R_{\bY_{t:(t-k)}}$ can be written as
\begin{multline}\label{eq:R_Y}
  R_{\bY_{t:(t-k)}|V_{t:(t-k)}=v_{t:(t-k)}} =
  \begin{pmatrix}
    \bSigma_{e_{S+1},e_{S+1},g_{S+1}} & \bOmega_{S}                     & & & \\
    \bOmega_{S}^\top   & \bSigma_{e_{S},e_{S},g_S} & \bOmega_{S-1} & & \\
                    & \bOmega_{S-1}^\top & \bSigma_{e_{S-1},e_{S-1},g_{S-2}} & \rotatebox{13}{$\ddots$} & \\
                    & & \rotatebox{13}{$\ddots$} & \rotatebox{13}{$\ddots$} & \bOmega_{1}\\
                    & & & \bOmega_{1}^\top & \bSigma_{e_{1},e_{1},g_1}
  \end{pmatrix},
\end{multline}
where blocks not shown consist of 0's.
With diagonals of the transition matrix $M_V$ close to 1, corresponding to rare regime switches, and small Markov order $k$, the main use of~\cref{eq:R_Y} would be with $S=1$; i.e., a single regime switch between times $t-k$ and $t$.

Notice that $R_{\bY_{t:(t-k)}|V_{t:(t-k)}=v_{t:(t-k)}}$ only depends on the realizations of latent regimes $v_t,\dots,v_{t-k}$, and positive definiteness of $R_{\bY_{t:(t-k)}|V_{t:(t-k)}=v_{t:(t-k)}}$ for any $(v_t,\dots,v_{t-k})\in \cbra{1,\dots,G}^{k+1}$ should be guaranteed through constraining the parameter set.
In the special case when $\rho_1=\cdots=\rho_d=0$, $\bOmega_1,\dots,\bOmega_S$ are all zero matrices and $R_{\bY_{t:(t-k)}|V_{t:(t-k)}=v_{t:(t-k)}}$ reduces to a diagonal block matrix, corresponding to the assumption that observations in different regime periods are independent.

An advantage of a margin-closed model is that the closure under margins property leads to a parsimonious dependence structure. It can be seen that the margin-closed regime-switching model with zero-mean Gaussian margins in each regime has only $G(kd+d(d+1)/2)+G(G-1)$ parameters. In comparison, the $d$-dimensional zero-mean Markov Switching Vector Autoregressive (MSVAR) model (see \cite{Cheng2016}, \cite{Sola1994}, and \cite{Hamilton1990}) with $G$ regimes has $G(kd^2+d(d+1)/2)+G(G-1)$ parameters.

The following example illustrates a case of a 2-dimensional regime-switching time series model with Markov order 3.

\begin{example}
\label{example:mchmm}

Consider the case of a bivariate time series model with two regimes.
Part~1 shows how to derive correlation matrices for values of the process within a given regime, whereas Part~2 focuses on the correlation structure for values of the process before and after a regime switch.
The matrices shown below have rounding to two decimal places.

\noindent\textbf{Part 1}.
For univariate subprocesses, suppose $k_{1,1}=k_{1,2}=1$ for variable 1
and $k_{2,1}=k_{2,2}=2$ for variable 2;
i.e, the Markov order of the two univariate subprocesses is 1 and 2, respectively, for both regimes.
From \cref{pro:markov_order}, the Markov order for the entire process is $k=1+\max_{i,g} k_{i,g}=3$.

Suppose the values of partial autocorrelations $\alpha_{i,\ell,g}$ and cross-sectional correlations $\rho_{i,j,g}$ between subprocesses are as follows:
\medskip
\\
\indent\textbf{parameters for regime 1}: $\alpha_{1,1,1}=0.8$;\ $\alpha_{2,1,1}=0.6, \alpha_{2,2,1}=0.5$;\ $\rho_{1,2,1}=0.7$;
\\
\indent\textbf{parameters for regime 2}: $\alpha_{1,1,2}=0.7$;\ $\alpha_{2,1,2}=0.4, \alpha_{2,2,2}=0.8$;\ $\rho_{1,2,2}=0.2$;
\\
\indent\textbf{parameters for both regimes at regime switching}: $\rho_1=0.25, \rho_2=0.35$.

\medskip

The partial autocorrelations for lags $\ell>k_{i,g}$ are 0; i.e., 
\[
  \alpha_{1,2,1}=\alpha_{1,3,1}=\alpha_{2,3,1}=0\quad \mbox{and}\quad
  \alpha_{1,2,2}=\alpha_{1,3,2}=\alpha_{2,3,2}=0.
\]

We first calculate correlation matrices $R_{1,g}$, $R_{2,g}$ and $R_{\bY,g}$, which are then used to derive $R_g$ for $g=1,2$.
Let $\bY_{i, t:(t-3)}=\rbra{Y_{i,t}, Y_{i,t-1}, Y_{i,t-2}, Y_{i,t-3}}^\top$
for $i=1,2$, and $\bY_{t:(t-3)}=\rbra{\bY_{t}^\top, \bY_{t-1}^\top,
\bY_{t-2}^\top, \bY_{t-3}^\top}^\top$.

As a Toeplitz correlation matrix of an AR(1) process, $R_{1,1}$ can be obtained from the partial autocorrelations $\alpha_{1,1,1}=0.8, \alpha_{1,2,1}=\alpha_{1,3,1}=0$, leading to:
\[
  R_{1,1} & = \cov\rbra{\bY_{1, t:(t-3)}, \bY_{1, t:(t-3)}\big| V_t=\cdots=V_{t-3}=1}=
  \bpmat
    1.00 &    0.80 &    0.64 &    0.51\\
    0.80 &    1.00 &    0.80 &    0.64\\
    0.64 &    0.80 &    1.00 &    0.80\\
    0.51 &    0.64 &    0.80 &    1.00
  \epmat.
\]
As a Toeplitz correlation matrix of an AR(2) process,  $R_{2,1}$ is obtained
from $\alpha_{2,1,1}=0.6$, $\alpha_{2,2,1}=0.5$, $\alpha_{2,3,1}=0$, leading to:
\[
  R_{2,1} & = \cov\rbra{\bY_{2, t:(t-3)}, \bY_{2, t:(t-3)}\big| V_t=\cdots=V_{t-3}=1}=
  \bpmat
    1.00 &    0.60 &    0.68 &    0.50\\
    0.60 &    1.00 &    0.60 &    0.68\\
    0.68 &    0.60 &    1.00 &    0.60\\
    0.50 &    0.68 &    0.60 &    1.00
  \epmat.
\]
The contemporaneous correlation matrix $R_{\bY,1}$ has only one parameter, $\rho_{1,2,1}=0.7$:
\[
  R_{\bY,1} = \cov\rbra{\bY_{t}|V_t=1} =
  \bpmat
    1 & \rho_{1,2,1}\\
    \rho_{1,2,1} & 1
  \epmat=
  \bpmat
    1.00 & 0.70\\
    0.70 & 1.00
  \epmat.
\]
Then, $R_1$, with inclusion of cross-correlations, can be derived using $R_{1,1}$ (in columns 1,3,5,7), $R_{2,1}$ (in columns 2,4,6,8), and
$R_{\bY,1}$ ($2\times2$ diagonal blocks):
\[
  R_1 & = \cov\rbra{\bY_{t:(t-3)}, \bY_{t:(t-3)}\big| V_t=\cdots=V_{t-3}=1} =
  \bpmat
    1.00 &    0.70 &    0.80 &    0.49 &    0.64 &    0.50 &    0.51 &    0.39\\
    0.70 &    1.00 &    0.56 &    0.60 &    0.45 &    0.68 &    0.36 &    0.50\\
    0.80 &    0.56 &    1.00 &    0.70 &    0.80 &    0.49 &    0.64 &    0.50\\
    0.49 &    0.60 &    0.70 &    1.00 &    0.56 &    0.60 &    0.45 &    0.68\\
    0.64 &    0.45 &    0.80 &    0.56 &    1.00 &    0.70 &    0.80 &    0.49\\
    0.50 &    0.68 &    0.49 &    0.60 &    0.70 &    1.00 &    0.56 &    0.60\\
    0.51 &    0.36 &    0.64 &    0.45 &    0.80 &    0.56 &    1.00 &    0.70\\
    0.39 &    0.50 &    0.50 &    0.68 &    0.49 &    0.60 &    0.70 &    1.00
  \epmat.
\]
The non-diagonal $2\times2$ blocks come from
the formulas in \cref{ap: mcvar} for margin-closed VAR(3) models.

For regime $g=2$, matrices $R_{1,2}$, $R_{2,2}$ and $R_{\bY,2}$ can be computed similarly and are given by
\[
& R_{1,2} = \cov\rbra{\bY_{1, t:(t-3)}, \bY_{1, t:(t-3)}\big| V_t=\cdots=V_{t-3}=2}=
\bpmat
  1.00 &    0.70 &    0.49 &    0.34\\
  0.70 &    1.00 &    0.70 &    0.49\\
  0.49 &    0.70 &    1.00 &    0.70\\
  0.34 &    0.49 &    0.70 &    1.00
\epmat,\\
& R_{2,2} = \cov\rbra{\bY_{2, t:(t-3)}, \bY_{2, t:(t-3)}\big| V_t=\cdots=V_{t-3}=2}=
\bpmat
  1.00 &    0.40 &    0.83 &    0.39\\
  0.40 &    1.00 &    0.40 &    0.83\\
  0.83 &    0.40 &    1.00 &    0.40\\
  0.39 &    0.83 &    0.40 &    1.00
\epmat,\\
& R_{\bY,2} = \cov\rbra{\bY_{t}|V_t=2} =
  \bpmat
    1 & \rho_{1,2,2}\\
    \rho_{1,2,2} & 1
  \epmat=
  \bpmat
    1.00 & 0.20\\
    0.20 & 1.00
  \epmat.
\]
Finally, $R_2$, with inclusion of cross-correlations, can be derived from $R_{1,2}, R_{2,2}$, and $R_{\bY,2}$:
\[
  R_2 & = \cov\rbra{\bY_{t:(t-3)}, \bY_{t:(t-3)}\big| V_t=\cdots=V_{t-3}=2}=
  \bpmat
    1.00 &    0.20 &    0.70 &    0.13 &    0.49 &    0.17 &    0.34 &    0.12\\
    0.20 &    1.00 &    0.14 &    0.40 &    0.10 &    0.83 &    0.07 &    0.39\\
    0.70 &    0.14 &    1.00 &    0.20 &    0.70 &    0.13 &    0.49 &    0.17\\
    0.13 &    0.40 &    0.20 &    1.00 &    0.14 &    0.40 &    0.10 &    0.83\\
    0.49 &    0.10 &    0.70 &    0.14 &    1.00 &    0.20 &    0.70 &    0.13\\
    0.17 &    0.83 &    0.13 &    0.40 &    0.20 &    1.00 &    0.14 &    0.40\\
    0.34 &    0.07 &    0.49 &    0.10 &    0.70 &    0.14 &    1.00 &    0.20\\
    0.12 &    0.39 &    0.17 &    0.83 &    0.13 &    0.40 &    0.20 &    1.00
  \epmat.
\]
Note that the feasibility of the parameter set can be verified by checking positive definiteness of $R_1$ and~$R_2$.

\noindent\textbf{Part 2}.
Next, some details are provided related to regime switching.
The parameters during regime-switching are specified by 
$P=\bsmat \rho_1 & 0\\ 0  & \rho_2 \esmat=
  \rbra{
  \begin{smallmatrix}
    0.25 & 0 \\ 0 & 0.35 \end{smallmatrix}}$.
Suppose $T=8$ and $v_1=v_2=v_3=v_4=1\neq v_5=v_6=v_7=v_8=2$ for
illustration of \cref{eq:R_Y}.
Matrix $R_{\bY_{t:(t-3)}}$ in \cref{eq:R_Y} can be obtained for $t=4,\dots,8$ based on $R_1$, $R_2$, $P$
and the values of $v_1,\dots,v_8$.
In this case, $S=1$ for period from $t=2$ to $t=5$ and $R_{\bY_{5:2}|V_{5:2}=v_{5:2}}$ can be written using \cref{eq:R_Y}:
\[
  R_{\bY_{5:2}|V_{5:2}=v_{5:2}} =
  \bpmat
  1.00 &    0.20 &    0.25 &    0.18 &    0.20 &    0.12 &    0.16 &    0.12\\
  0.20 &    1.00 &    0.24 &    0.35 &    0.20 &    0.21 &    0.16 &    0.24\\
  0.25 &    0.24 &    1.00 &    0.70 &    0.80 &    0.49 &    0.64 &    0.50\\
  0.18 &    0.35 &    0.70 &    1.00 &    0.56 &    0.60 &    0.45 &    0.68\\
  0.20 &    0.20 &    0.80 &    0.56 &    1.00 &    0.70 &    0.80 &    0.49\\
  0.12 &    0.21 &    0.49 &    0.60 &    0.70 &    1.00 &    0.56 &    0.60\\
  0.16 &    0.16 &    0.64 &    0.45 &    0.80 &    0.56 &    1.00 &    0.70\\
  0.12 &    0.24 &    0.50 &    0.68 &    0.49 &    0.60 &    0.70 &    1.00
  \epmat.
\]
In the above,
\begin{itemize}
\itemsep=0pt
\item
$\cov(\bY_5,\bY_4)==P\corr(\bY_4)=P\cov(\bY_4,\bY_4)$ comes from \cref{eq:condition1}.
\item
$\bY_5\perp\bY_3\mid\bY_4$ in \cref{eq:condition2} implies
$\bm{0}=\cov(\bY_5,\bY_3) - \cov(\bY_5,\bY_4)[\cov(\bY_4,\bY_4)]^{-1}
\cov(\bY_4,\bY_3)$ so that
$\cov(\bY_5,\bY_3) = P\cov(\bY_4,\bY_4) [\cov(\bY_4,\bY_4)]^{-1}
\cov(\bY_4,\bY_3) = P\cov(\bY_4,\bY_3)$.
\item
$\bY_5\perp\bY_2\mid\bY_4$ in \cref{eq:condition2} similarly implies
$\cov(\bY_5,\bY_2) = P\cov(\bY_4,\bY_2)$.
\end{itemize}

Since $\cbra{\bY_t^\top, \dots, \bY_{t-3}^\top}^\top$ follows a multivariate Gaussian distribution with mean zero and covariance matrix $R_{\bY_{t:(t-3)}}$,
the following stochastic representations can be derived
from conditional Gaussian distributions (based on the correlation
matrix of $\bY_{8:2}$ using  \cref{eq:condition1}--\cref{eq:condition3}):
\[
&\bY_4 =
\bsmat
  1.11 &  -0.33\\
  0.72 &  -0.01
\esmat
\bY_3 +
\bsmat
  -0.33 &    0.38\\
  -0.70 &    0.82
\esmat
\bY_2 +
\bepsilon_4,
&\bepsilon_4\sim \distNorm_2\rbra{\bm{0},
\bsmat
  0.30 &    0.17\\
  0.17 &    0.35
  \esmat
};\\
&\bY_5 =
\bsmat
  0.25 &    0.00\\
  0.00 &    0.35
  \esmat
\bY_4 + \bepsilon_5,\
&\bepsilon_5\sim \distNorm_2\rbra{\bm{0},
\bsmat
  0.94 &    0.14\\
  0.14 &    0.88
  \esmat
};\\
&\bY_6 =
\bsmat
  0.74 &    0.03\\
  0.08 &    0.44
  \esmat
\bY_5 +
\bsmat
  -0.19 &  -0.01\\
  -0.02 &  -0.15
  \esmat
\bY_4 +
\bepsilon_6,
&\bepsilon_6\sim \distNorm_2\rbra{\bm{0},
\bsmat
  0.48 &    0.08\\
  0.08 &    0.81
  \esmat
};\\
&\bY_7 =
\bsmat
  0.71 &  -0.06\\
  0.09 &    0.02
  \esmat
\bY_6 +
\bsmat
  -0.01 &    0.12\\
  -0.10 &    0.94
  \esmat
\bY_5 +\bsmat
  0.00 &  -0.04\\
  0.03 &  -0.33
  \esmat
\bY_4 +
\bepsilon_7,&\bepsilon_7\sim \distNorm_2\rbra{\bm{0},
\bsmat
  0.50 &    0.03\\
  0.03 &    0.21
\esmat
};\\
&\bY_8 =
\bsmat
  0.71 &  -0.05\\
  0.15 &    0.07
\esmat
\bY_7 +
\bsmat
  -0.02 &    0.10\\
  -0.18 &    0.82
\esmat
\bY_6 +
\bepsilon_8,
&\bepsilon_8\sim \distNorm_2\rbra{\bm{0},
\bsmat
  0.50 &    0.04\\
  0.04 &    0.29
\esmat
}.
\]
Note that $\bY_1$ and $\bY_5$ do not appear in the first and last equations above, respectively, because the Markov order is $k-1=2$ within a fixed latent regime.

In the special case that $\rho_1=\rho_2=0$ (independence between regimes), $\bY_1,\dots,\bY_4$ are independent of $\bY_5,\dots,\bY_8$. Matrices $R_{\bY_{5:2}}, R_{\bY_{6:3}}, R_{\bY_{7:4}}$ are block diagonal, and the stochastic representations for $\bY_5, \bY_6$, and $\bY_7$ become
\[
&\bY_5 = \bepsilon_5,\
&\bepsilon_5\sim \distNorm_2\rbra{\bm{0},
\bsmat
  1.00 &    0.20\\
  0.20 &    1.00
\esmat
};\\
&\bY_6 =
\bsmat
  0.70 &    -0.01\\
  0.06 &    0.39
\esmat
\bY_5 +
\bepsilon_6,
&\bepsilon_6\sim \distNorm_2\rbra{\bm{0},
\bsmat
  0.51 &    0.11\\
  0.11 &    0.84
\esmat
};\\
&\bY_7 =
\bsmat
  0.71 &  -0.05\\
  0.15 &    0.07
\esmat
\bY_6 +
\bsmat
  -0.02 &    0.10\\
  -0.18 &    0.82
\esmat
\bY_5 + \bepsilon_7,&\bepsilon_7\sim \distNorm_2\rbra{\bm{0},
\bsmat
  0.50 &    0.04\\
  0.04 &    0.29
\esmat
}.
\]
The main difference is the omission of $\bY_4$ in the above equations; this is consistent with observations before and after regime change being independent.
\end{example}

\section{Inference}
\label{sec:fitting_models}

This section discusses several approaches to estimate parameters of the regime switching model
introduced in \cref{def:model} as well as to make inference for the latent regime sequence.
These approaches take advantage of the property of closure under margins.
We give the expressions of joint density functions of the observations.
Then, \cref{sec:with_external_info} considers the case when the external information on the latent regime sequence is available, and \cref{sec:without_external_info} explores the general situation where the latent regime sequence is inferred from the observations.


We assume an observed multivariate time series of length $T$ is
$(\bx_1,\dots,\bx_T)$ where $\bx_t=\rbra{x_{1,t},\dots,x_{d,t}}^\top$
is a realization of $\bX_t$ for $1\leq t\leq T$.
To consider the model in \cref{def:model} after probability integral transforms, we assume that exploratory data analysis suggests a few regimes
based on shifts in location and/or scatter.


With parametric families for the univariate distributions,
we now write
$F_{i,g}=F_{i,g}\rbra{\cdot;\boldeta_{i,g}}$ for an absolutely
continuous parametric family with parameter $\boldeta_{i,g}$ for the $i$-th marginal component in regime~$g$.
The parametric families would be chosen to
handle skewness and tail behavior seen in the observed data.
\cref{eq:PIT} becomes
 $$Y_{i,t}= a_{i,g}\rbra{X_{i,t};\boldeta_{i,g}}, \quad
 1\leq i\leq d, \quad V_t=g,$$
and the derivative of the transform is
\begin{equation}
  a'_{i,g}\rbra{x;\boldeta_{i,g}} = f_{i,g}(x;\boldeta_{i,g})
  \bigm/\phi\circ\Phi^{-1}\bigl(F_{i,g}(x;\boldeta_{i,g}) \bigr),
  \label{eq:transform-derivative}
\end{equation}
where $\phi$ is the $\distNorm(0,1)$ density and $f_{i,g}=F'_{i,g}$.
The regime-switching model in \cref{def:model} applies
to the transformed multivariate time series $\{\bY_t\}$.


We consider
two approaches to determining the Markov order $k$.
In the case when long stationary segments exist for all regimes and one can roughly distinguish them, AR models can be fitted for each transformed univariate series in each regime after all univariate margins are fitted.
Then, $k$ can be taken to be 1 plus the maximum Markov order of these AR models.
Otherwise, the Akaike information criterion (AIC) can be adopted as the criterion for Markov order determination.

One group of hyperparameters and five groups of parameters are itemized
as follows.
\begin{enumerate}
\itemsep=0pt
\item[0.] Markov order hyperparameters $k_{i,g}$ for $1\leq i\leq d$ and $1\leq g\leq G$.
\item[1.] $d$ univariate marginal distributions for each of $G$ regimes:
$\btheta_{\boldeta}=\{\boldeta_{1,1},\dots,\boldeta_{1,G},\dots,\boldeta_{d,1},\dots,\boldeta_{d,G}\}$.
\item[2.] Toeplitz correlation matrices of serial dependence for each of $d$ variables and $G$ regimes;
$\btheta_R=\{R_{1,1},\dots,R_{1,G},\dots,R_{d,1},\dots,R_{d,G}\}$.
\item[3.] Contemporaneous correlation matrices for each of $G$ regimes:
$\btheta_{R_{\bY}}=\cbra{R_{\bY,g}: 1\leq g\leq G}$.
\item[4.] Serial correlations during regime switching:
$\btheta_{P}=P$ which is a diagonal $d\times d$ matrix.
\item[5.] Dynamics of the hidden Markov chain: $\btheta_{V}=\cbra{\bp_V,
M_V}$, where $\bp_V$ is the probability of initial state and $M_V$ is the transition matrix.
\end{enumerate}

Note that cross-correlations of the $\bY_t$ and $\bY_{t-\ell}$ in the same
regime can be derived from items 2 and 3 with the margin-closed VAR
assumption within each regime.

Benefiting from the property of closure under margins, the serial correlations of each univariate component can be estimated separately, i.e., the cross-sectional correlations can be ignored when serial correlations are being estimated.
Specifically, for each regime $g$ in $\cbra{1,\dots,G}$, the parameters of
regime $g$ can be estimated using the data segments in the regime through a multi-stage procedure,
in which the parameters of univariate components (items 1 and 2) are estimated first,
followed by estimation of cross-sectional parameters (item 3).
The sequential estimation follows some of the steps in Section 4 of \cite{Zhang2023} for margin-closed VAR models and Section 5.5 of \cite{Joe2014} for copula-based models.

Let $\btheta$ denote the set of all parameters.
Let $\phi_{D}(\cdot;\bSigma)$ denote the $D$-variate multivariate
Gaussian density with mean vector \textbf{0} and covariance matrix
$\bSigma$.
Let $y_{i,t-\ell}=a_{i,v_{t-\ell}}\rbra{x_{i,{t-\ell}};\boldeta_{i,v_{t-\ell}}}$ for $0\leq \ell\leq k$, $1\leq i\leq d$, and $t=\ell+1,...,T$.
As \cref{pro:markov_order} indicates Markov order $k$ for the time
series, likelihood calculations (given in \cref{sec:with_external_info} below) require
the joint density of $k+1$ consecutive observations, say from time $t-k$ to $t$, given $V_t=v_t,\dots, V_{t-k}=v_{t-k}$:
\begin{equation}\label{eq:multi_density}
  \begin{split}
    &f_{\bX_{t:(t-k)}|V_{t:(t-k)}}(\bx_t,\dots,\bx_{t-k}|v_t,\dots,v_{t-k}; \btheta)\\
   & \quad= \phi_{(k+1)d}\rbra{y_{1,t},\dots,y_{d,t},\dots,y_{1,t-k},\dots,y_{d,t-k};   R_{\bY_{t:(t-k)}|V_{t:(t-k)}=v_{t:(t-k)}}}\\
     & \qquad\qquad\times \prod_{\ell=0}^k\prod_{i=1}^d a'_{i, v_{t-\ell}}\rbra{x_{i,t-\ell}; \boldeta_{i, v_{t-\ell}}},\quad\ t=k+1,\dots,T,
  \end{split}
\end{equation}
where $a'_{i,g}$ is defined in \cref{eq:transform-derivative}


If $v_t=\cdots=v_{t-k}=g$ for a specific regime $g$, i.e., there is no regime switching from $t-k$ to $t$, \cref{eq:multi_density} can be simplified by setting
$R_{\bY_{t:(t-k)}|V_{t:(t-k)}=v_{t:(t-k)}}=R_g$, leading to the conditional densities of the form:
\begin{equation}\label{eq:multi_density_of_const_reg}
  \begin{split}
    &f_{\bX_{t:(t-k)}|V_{t:(t-k)}}(\bx_t,\dots,\bx_{t-k}|g,\dots,g; R_{\bY,g}, \boldeta_{i, g},R_{i,g}, 1\leq i\leq d)\\
   & \quad = \phi_{(k+1)d}\rbra{y_{1,t},\dots,y_{d,t},\dots,y_{1,t-k},\dots,y_{d,t-k}; R_g}\times
  \prod_{\ell=0}^k\prod_{i=1}^d a'_{i, g}\rbra{x_{i,t-\ell}; \boldeta_{i, g}},\quad t=k+1,\dots,T.
  \end{split}
\end{equation}
Furthermore, when $v_t=\cdots=v_{t-k}=g$ and if only the density
of the $i$-th univariate component is used to estimate parameters
$\boldeta_{i, g}$ and $R_{i,g}$,
the univariate version of \cref{eq:multi_density_of_const_reg}:
\begin{equation}\label{eq:uni_density_of_const_reg}
  \begin{split}
    &f_{X_{i,t:(t-k)}|V_{t:(t-k)}}(x_{i,t},\dots,x_{i,t-k}|g,\dots,g; \boldeta_{i, g},R_{i,g})\\
    & \quad = \phi_{k+1}\rbra{y_{i,t},\dots,y_{i,t-k}; R_{i,g}}\times
  \prod_{\ell=0}^k a'_{i, g}\rbra{x_{i,t-\ell}; \boldeta_{i, g}}
  \end{split}
\end{equation}
can be used.
The estimation of parameters can be done based on \eqref{eq:multi_density} -- \eqref{eq:uni_density_of_const_reg}.

\subsection{Estimation with external information on regimes}
\label{sec:with_external_info}

In this section, it is explained how the model can be fitted using external information on the latent regimes, i.e., when the latent regime sequence is given.
\cref{sec:estimate_with_external} provides details on estimation of the model parameters given a latent regime sequence.
Then, \cref{sec:latent_infer_with_external} introduces updating of the latent regime sequence by combining the statistical model with the externally given latent regime sequence.

There may be cases where external information on the latent regime switching is available.
A typical example includes a multivariate time series of macroeconomic indicators, where regimes correspond to business cycles.

With external information on regimes,
the original time series can be partitioned into several long contiguous
segments split by the times of regime switches.
Each segment has observations in a single regime.
We start with the same notation as in derivation of \cref{eq:R_Y} but extend it to the whole series $V_1=v_1,\dots,V_T=v_T$.
Let the number of regime switches be $S$.
Then the sequence $v_1,\dots,v_T$ can be partitioned into $S+1$ segments.
Let $t_1,\dots,t_S$ be the time points of regime switches with regime values specified as:
\begin{equation}
  \begin{gathered}
    v_{t_S}=\cdots=v_T=g_{S+1},\\
    v_{t_{S-1}}=\cdots=v_{t_S-1}=g_S \neq v_{t_S},\\
    \vdots\\
    v_{1}=\cdots= v_{t_1-1} =g_1 \neq v_{t_1}.
  \end{gathered}
\end{equation}
Let $t_0=1$ and suppose $t_{S+1}=T+1$.
The $s$-th segment of the regime sequence is denoted by $V_{t_{s-1}:(t_s-1)}$ for $1\leq s \leq S+1$ \blue{(you seem to use boldface notion $\bV$ below, not consistently though)}.
Then, based on \cref{eq:multi_density}, the log-likelihood of the multivariate time series given the external latent regime sequence is given by
\begin{equation}\label{eq:likelihood_given_latent}
  \begin{split}
     &\ell_{\bX_{1:T}|V_{1:T}}\rbra{P, R_{\bY, g}, R_{i, g}, \boldeta_{i, g},1\leq i \leq d, 1\leq g \leq G \big|\bx_{1},\dots,\bx_{T},v_1,\dots,v_T}\\
     = & \sum_{t=k+2}^{T} \log f_{\bX_{t}|\bX_{(t-1):(t-k)},\bV_{t:(t-k)}}(\bx_t|\bx_{t-1},\dots,\bx_{t-k},v_t,\dots,v_{t-k}; \btheta) \\
     & \hspace{4cm} + \log f_{\bX_{1:(k+1)}|\bV_{1:(k+1)}}(\bx_1,\dots,\bx_{k+1}|v_1,\dots,v_{k+1}; \btheta).
  \end{split}
\end{equation}
The conditional density $f_{\bX_{t}|\bX_{(t-1):(t-k)},\bV_{t:(t-k)}}$ can be derived analytically from \cref{eq:multi_density} and conditional distributions of multivariate Gaussian random vectors.
The log-likelihood of the multivariate $s$-th segment given
$\{V_t\}_{t=1,\dots,T}$ is 
\begin{equation}\label{eq:multi_loglik_with_external}
\begin{split}
   &\ell_{s|V_{1:T}}\rbra{R_{\bY,g_s}, R_{i, g_s}, \boldeta_{i, g_s},1\leq i \leq d \big|\bx_1,\dots,\bx_T,v_1,\dots,v_T} =\\
   &\left\{
  \begin{aligned}
     & \log f_{\bX_{(t_s-1):t_{s-1}}|V_{(t_s-1):t_{s-1}}}(\bx_{t_s-1},\dots,\bx_{t_{s-1}}|g_s,\dots,g_s; R_{\bY,g_s}, \boldeta_{i, g_s},R_{i,g_s}, 1\leq i\leq d),\\
     &\hspace{10cm} \mbox{if}\ t_{s}-t_{s-1}\leq k+1  \\
     & \sum_{t=t_{s-1}+k+1}^{t_s-1}
 \log f_{\bX_t|\bX_{(t-1):(t-k)}, V_{t:(t-k)}}(\bx_t|\bx_{t-1},\dots,\bx_{t-k},g_s,\dots,g_s;R_{\bY,g_s},  \boldeta_{i, g_s},R_{i,g_s},1\leq i\leq d) \\
     &\hspace{1cm}+\log
  f_{\bX_{(t_{s-1}+k):t_{s-1}}|V_{(t_{s-1}+k):t_{s-1}}}(\bx_{t_{s-1}+k},
  \dots,\bx_{t_{s-1}}|
     g_s,\dots,g_s; \boldeta_{i, g_s}, R_{i,g_s}, R_{\bY,g_s},\\
     & \hspace{10cm} 1\leq i\leq d),\ \mbox{otherwise},
  \end{aligned}
  \right.
\end{split}
\end{equation}
where the conditional density $f_{\bX_t|\bX_{(t-1):(t-k)}, V_{t:(t-k)}}$ can be analytical derived from \cref{eq:multi_density_of_const_reg}.
The log-likelihood of the $i$-th univariate component in the $s$-th
segments given the external latent regime sequence can be obtained based on \cref{eq:uni_density_of_const_reg}:
\begin{equation}\label{eq:uni_loglik_with_external}
\begin{split}
   &\ell_{i,s|V_{1:T}}\rbra{R_{i,g_s}, \boldeta_{i,g_s}\big|x_{i,1},\dots,x_{i,T},v_1,\dots,v_T} =\\
   &\left\{
  \begin{aligned}
     & \log f_{X_{(t_s-1):t_{s-1}}|V_{(t_s-1):t_{s-1}}}(x_{i,t_s-1},\dots,x_{i,t_{s-1}}|g_s,\dots,g_s; \boldeta_{i, g_s},R_{i,g_s}),\ \mbox{if}\ t_{s}-t_{s-1}\leq k+1;\\
     & \sum_{t=t_{s-1}+k}^{t_s-1}
 \log f_{X_{i,t}|X_{i,(t-1):(t-k)}|V_{t:(t-k)}}(x_{i,t}|x_{i,t-1},\dots,x_{i,t-k}|
   g_s,\dots,g_s; \boldeta_{i, g_s},R_{i,g_s})\\
     & \hspace{1cm}+\log
f_{X_{i,(t_{s-1}+k):t_{s-1}}|V_{(t_{s-1}+k):t_{s-1}}}(x_{i,t_{s-1}+k},
\dots,x_{i,t_{s-1}}|g_s,\dots,g_s; \boldeta_{i, g_s},R_{i,g_s}),\ \mbox{otherwise}.
  \end{aligned}
  \right.
\end{split}
\end{equation}
If the actual time points of regime switches are off by 1 or 2 time
units and the sojourn time in each state is long enough,
misspecification of the points of regime switches will have little effect
on the parameter estimation.

\subsubsection{Parameter estimation given a latent regime sequence}
\label{sec:estimate_with_external}

The parameters of the hidden Markov chain can be estimated by maximizing the likelihood of the regime sequence, i.e.,
 $$\hat{\bp}_V=\rbra{\bm{1}_{\cbra{v_1=1}},\dots,\bm{1}_{\cbra{v_1=G}}}^\top,
 \quad
 \widehat{M}_{V, g,g'}=\frac{\sum_{s=1}^{S}\bm{1}_{\cbra{v_{t_{s}-1}=g,
v_{t_s}=g'}}}{\sum_{s=1}^{S}\bm{1}_{\cbra{v_{t_{s-1}}=g}}},$$
where $\bm{1}_{\cbra{\cdot}}$ is the indicator function, and $\widehat{M}_{V, g,g'}$ is the estimate of the element in row $g$ and column $g'$ of $M_V$
based on the time points of regime switches.
Then a sequential estimation procedure can be performed with the following four steps.
In the first step, parameters  of the univariate marginal distributions are estimated with the
serial and cross-sectional dependence ignored.
In the second step, parameters of the univariate margins are fixed at the estimates obtained 
in Step~1, and the serial correlations during regime switching as well as
the cross-sectional correlations are ignored when computing the likelihood
of all segments.

\medskip

\indent \textbf{Step 1}.
For  $1\leq i \leq d$ and $1\leq g\leq G$,
estimate the univariate margin parameters $\boldeta_{i,g}$
by maximizing the quasi-likelihood given $v_1,\dots,v_T$,
ignoring serial dependence:
\[\tilde{\boldeta}_{i,g} =
\arg\max_{\boldeta_{i,g}}\sum_{\substack{v_{t_s-1}=g\\1\leq s\leq S+1}}
  \  \sum_{t=t_{s-1}}^{t_s-1}\log f_{i, g}\rbra{x_{i,t};\boldeta_{i,g}}.\]

\indent  \textbf{Step 2}.
For  $1\leq i \leq d$ and $1\leq g\leq G$,
estimate the Toeplitz correlation matrices $R_{i,g}$
by maximizing the likelihood of each univariate component in each regime
based on \cref{eq:uni_loglik_with_external}:
\[\tdR_{i,g}=\arg\max_{R_{i,g}}\sum_{\substack{v_{t_s-1}=g\\1\leq s\leq
S+1}}\ell_{i,s|V_{1:T}}\rbra{R_{i,g_s},
\hat{\boldeta}_{i,g_s}\big|x_{i,1},\dots,x_{i,T},v_1,\dots,v_T}.\]

After going through the two steps above for all regimes, we next estimate the serial correlations during regime switching and the cross-sectional correlations between univariate components.
In Step 3, the estimates of univariate margins and Toeplitz correlation matrices in all regimes are fixed at the estimated values in previous steps.
In Step 4, all parameters are fixed except for $P$.


\medskip

\indent \textbf{Step 3}.
For $1\leq g\leq G$,
estimate the cross-sectional correlation matrix $R_{\bY,g}$
by maximizing the likelihood of multivariate segments
based on \cref{eq:multi_loglik_with_external}:
\[\tdR_{\bY,g}=\arg\max_{R_{\bY,g}}\sum_{\substack{v_{t_s-1}=g\\1\leq s\leq
S+1}}\ell_{s|V_{1:T}}\rbra{R_{\bY,g_s}, \hat{R}_{i, g_s}, \hat{\boldeta}_{i,
g_s}, 1\leq i \leq d\Big|\bx_{t_{s-1}},\dots,\bx_{t_s-1},v_1,\dots,v_T}.\]

\indent \textbf{Step 4}. Estimate the serial correlations parameters during regime switching: $P = \diag\rbra{\rho_1, \dots, \rho_d}$ by maximizing the likelihood of the original multivariate series given the  external latent regime sequence, i.e.,
\[\tdP=\arg\max_{P}\ell_{\bX_{1:T}|V_{1:T}}\rbra{P, \hat{R}_{\bY,g}, \hat{R}_{i, g}, \hat{\boldeta}_{i, g}, 1\leq i \leq d,  1\leq g \leq G\Big|\bx_{1},\dots,\bx_{T},v_1,\dots,v_T}.\]

\subsubsection{Regime sequence updating}
\label{sec:latent_infer_with_external}

After estimating model parameters based on external information on the regimes, inferring the latent regime sequence based on the obtained estimates is also meaningful.
The idea is to adjust the latent regime sequence based on external
information by incorporating the statistical model.
The setting of a low probability of regime switching enables us to only infer the times of regime switching.
Let $p_{t, \tau}(g)$ be the probability that $V_{t}=\cdots=V_{t+\tau}=g$, given the observed time series and estimated parameters.
The algorithm of computing $p_{t, \tau}(g)$ is provided in \cref{ap: baum-welch}.
We then determine the updated regime sequence based on the following rule.

First, the initial regime $v_0$ is determined by $v_0 = \arg\max_{g\in\{1,\dots,G\}}p_{0, 0}(g)$.
Then, to detect a regime switching, we require the probability of $p_{t, \tau}(g')$ for a different regime $g'$ to exceed a threshold probability value $\xi$ for consecutive time steps of length $\nu$.
That is,
the current regime $g$ stays until the time of the next regime switching,
denoted by $t_{switch}$, which is ``detected" through the condition that
$\min\cbra{p_{t_{switch}+i, \tau}(g'): i=0,1,\dots,\nu}>\xi$ for a different regime $g'\neq g$.

The parameter $\tau$ is considered as a smoothing parameter as
a larger $\tau$ indicates a smoother function $p_{t, \tau}(g)$ with respect to $t$.
When $\tau=0$, $p_{t, \tau}(g)$ is the conditional probability of being in regime $g$ at  time $t$ given the observations and parameter estimates.
\cite{chauvet2008comparison} use it to determine the latent turning point dates of business cycles  based on macroeconomic indicators with $\xi=0.8$ and $\nu=3$.
But the conditional probability for $\tau=0$ is usually volatile, and a smoother alternative of the conditional probability for a period of time points is preferred when only a limited number of regime switches is desired.

An idea combining the external information on the latent regime sequence and the statistical model is to replace the external latent regime sequence with the updated regime sequence and estimate all parameters by following Step 1--4 in \cref{sec:estimate_with_external} again.
Consequently, the regime sequence can be updated again with the new estimated parameters.
It can be performed repeatedly until there is no difference before and after
the regime sequence updating.

\subsection{Estimation using observed time series only}
\label{sec:without_external_info}

In this section, a procedure is given for estimating parameters of the
regime-switching model without external information on the regime sequence.
\cref{sec:multi-stage_approach} sketches a multi-stage estimation procedure, and \cref{sec:iterative_approach} gives an iterative estimation procedure based on the inferred latent regime sequence.

Two special techniques taking advantage of the closure under margins, which implies that any subprocess of the multivariate
time series is a regime-switching model
with the same Markov order $k$, the same latent regime sequence,
the same $\btheta_V$
and other parameters that are subsetted from
$\btheta_{\boldeta}, \btheta_R, \btheta_{R_{\bY}}, \btheta_P$, can be applied.

First of all, as all marginal sub-processes of the closed-under-margins regime-switching process are also regime-switching processes sharing the same latent regime sequence, the inference for the latent regime sequence can be made based on only a subset of the components of the multivariate time series.
Theoretically, using more components should lead to more accurate and reliable estimates of the latent regime sequence.
But different components may contribute differently to the ability to determine latent regimes.
A subset of the components may be adequate to infer the latent regime sequence, based on which the parameters of the remaining components can be estimated following the procedure in \cref{sec:with_external_info}.
Apart from the expert knowledge on which components are more important and useful, a statistical idea is to select those components that have large distances between their univariate marginal distributions in different regimes.
A simulation study in \cref{sec:sim_subset} supports this idea.

Then, with the subset of univariate components selected, a special fitting procedure can be applied to the model.
In the following subsections, two approaches are proposed to obtain the estimates of parameters based on the observed time series.
The first method is to maximize the likelihood through a multi-stage procedure by utilizing the property of closure under margins.
For the second method, we follow a similar idea as when the external information is available: the parameters are iteratively updated until the inferred latent regime sequence is stable.

\subsubsection{Multi-stage estimation procedure}
\label{sec:multi-stage_approach}

As we try to obtain the maximum likelihood estimates of all parameters, we still suggest turning to a multi-stage procedure, which takes advantage of closure under margins.
The notation for different groups of model parameters is summarized at the
beginning of \cref{sec:fitting_models}.
As the log-likelihood of observations given a latent regime sequence is provided by \cref{eq:likelihood_given_latent}, the likelihood of observations can be expressed by marginalizing out the latent regime sequence:
\begin{equation}\label{eq:likelihood_of_obs}
  \begin{split}
    & \ell_{\bX_{1:T}}\rbra{\btheta_{\boldeta}, \btheta_R, \btheta_{R_{\bY}}, \btheta_{P}, \btheta_M\big|\bx_{1:T}} \\
    = &\log\rbra{\sum_{v_{1:T}\in\cbra{1,\dots,G}^T}\exp\cbra{\ell_{\bX_{1:T}|V_{1:T}=v_{1:T}}\rbra{\btheta_{\boldeta}, \btheta_R, \btheta_{R_{\bY}}, \btheta_{P}\big|\bx_{1:T}}}p_{V_{1:T}}\rbra{v_1,\dots,v_T|\btheta_M}},
  \end{split}
\end{equation}
where $p_{V_{1:T}}\rbra{v_1,\dots,v_T|\btheta_M}$ is the probability that $V_1=v_1,\dots,V_T=v_T$ given $\bp_V$ and $M_V$. It is given by
\[
  p_{V_{1:T}}\rbra{v_1,\dots,v_T|\btheta_M} = \bp_{V,v_1}\prod_{t=2}^{T} M_{V, v_{t-1}, v_t},
\]
where $\bp_{V,g}=\Pr(V=g)$ and $M_{V, v_{t-1}, v_t}$ is the (one-step) transition
probability from $v_{t-1}$ to $v_t$.
The calculation of \cref{eq:likelihood_of_obs} can be performed by a generalized Baum-Welch Algorithm; see details in \cref{ap: baum-welch}.

\medskip

\textbf{Step 1}. Estimate univariate margin parameters $\btheta_{\boldeta}$ and parameters of the hidden Markov chain $\btheta_M$ while ignoring all correlation parameters.
It is equivalent to fitting a hidden Markov model with independent univariate random variables in emission distributions; see \cite{HMM2017}.

\textbf{Step 2}. Fix parameters $\btheta_{\boldeta}$ and $\btheta_M$ at their estimates in Step 1, and estimate the set of Toeplitz correlation matrices $\btheta_R$ with the cross-sectional correlations and serial correlations during regime switching ignored, through maximizing the objective function in \cref{eq:likelihood_of_obs}.

\textbf{Step 3}. Fix parameters $\btheta_{\boldeta}$, $\btheta_M$, and $\btheta_R$ at their estimates in Steps 1--2, and estimate the cross-sectional correlation matrix $\btheta_{R_{\bY}}$ with the serial correlations during regime switching ignored, through maximizing the objective function in \cref{eq:likelihood_of_obs}.

\textbf{Step 4}. Fix parameters estimated in Steps 1--3, and estimate the set of serial correlation parameters during regime switching, i.e., estimate $\btheta_{P}$ by maximizing the objective function in \cref{eq:likelihood_of_obs}.

Note that Steps 3 and 4 are required to be performed under the positive definiteness constraint of matrices in \cref{eq:R_Y}.

\subsubsection{Iterative estimation based on the inferred latent regime sequence}
\label{sec:iterative_approach}

A problem with maximizing the likelihood of multivariate observations is a high computational cost of the constrained optimizations, especially in the case of a high-dimensional observed time series.
In contrast, maximizing the ``complete" likelihood as in \cref{eq:likelihood_given_latent} is much simpler.
Therefore, an alternative approach similar to the method in \cref{sec:estimate_with_external} is proposed here.

First, an initial value of the latent regime sequence is calculated. To do this, we suggest fitting a hidden Markov model for independent univariate random variables in emission distributions to obtain initial estimates for the marginal parameters $\boldeta_{1,g},\dots,\boldeta_{d,g}\ \mbox{for}\ g \in \cbra{1,\dots,G}$ and hidden Markov chain parameters $\bp_V, M_V$, and setting $R_{i,g}$ and $R_{\bY,g}$ as $(k+1)\times (k+1)$ and $d\times d$ identity matrices, respectively, for $1\leq i\leq d, 1\leq g\leq G$, and $P$ as a $d\times d$ zero matrix.
With these initial model parameter estimates, an initial latent regime sequence can be inferred using the method in \cref{sec:latent_infer_with_external}. Then, the parameter estimates are updated by maximizing the complete likelihood in \cref{eq:likelihood_given_latent}, repeating the estimation steps in \cref{sec:latent_infer_with_external}.
The procedure is run iteratively until the inferred latent regime sequence becomes stable.

\section{Simulation Study}
\label{sec:simulation}

In this section, we illustrate and validate the proposed estimation procedures
with some simulated data sets based on the
margin-closed hidden Markov model in \cref{def:model}.
In \cref{sec:sim_fit_methods}, the simulated data is fitted using external
information on the latent regime sequence, and the estimates are investigated.
\cref{sec:sim_subset} studies the effects of variable subset selection on latent regime inference based on the simulated data.

A 4-dimensional time series of length $1,000$ with two latent regimes is generated in each simulation.
The Markov order of all univariate components in both regimes is set to 1. That is, $d=4$, $G=2$, $k_{i,g}=1$ for all $i=1,\dots,4$ and $g=1,2$. The marginal components in all regimes follow the
skew-t distribution \citep{Jones2003skew_t} which is characterized by location, scale,
left tailweight and right tailweight parameters.
The tailweight parameters correspond to the index of regular variation for
the tails of the distribution, so that larger tailweight parameters
indicate a lighter tail, closer to Gaussian exponentially decaying tails.
The adopted margin parameters and the parameters of the latent Markov chain in the simulation are
\[
  &\boldeta_{1,1} = \rbra{0.0, 1.0, 4.0, 8.0}^\top, \boldeta_{1,2} =
\rbra{4.0, 1.0, 4.0, 8.0}^\top; \\
  &\boldeta_{2,1} = \rbra{0.0, 1.0, 4.0, 8.0}^\top, \boldeta_{2,2} =
\rbra{2.0, 1.0, 4.0, 8.0}^\top; \\
  &\boldeta_{3,1} = \rbra{1.0, 2.0, 4.0, 8.0}^\top, \boldeta_{4,2} =
\rbra{1.0, 2.0, 4.0, 8.0}^\top; \\
  &\boldeta_{4,1} = \rbra{0.0, 2.0, 4.0, 8.0}^\top, \boldeta_{4,2} =
\rbra{0.0, 2.0, 4.0, 8.0}^\top; \\
  &\bp_V = \rbra{0.5, 0.5}^\top,
  M_V = \bsmat
  0.95 &    0.05\\
  0.02 &    0.98
  \esmat;
\]
where the elements in univariate parameter vector $\boldeta_{i,g}$ are, in sequence, the location, scale, left tailweight, and right tailweight parameters.
Other parameters in the simulation can be found in Tables \ref{tb:true_and_estimated_within_regime} and \ref{tb:true_and_estimated_regime_switching}.

The univariate parameter vectors are chosen so that for different
components, the distances between univariate distributions in two regimes
can be directly compared, i.e., all margins have the same left and right
tailweight parameters:
$4$ and $8$, respectively, so that all variables are skewed and heavy-tailed.
For the marginal sub-process comprising components 1 and 4, and the sub-process comprising components 2 and 4, both their cross-sectional correlations shift from $0.2$ to $0.1$ when the regime switches from 1 to 2, but the univariate component 1 has larger distance between univariate margins in two regimes than the univariate component 2.
The sub-process of components 3 and 4 has regime-invariant univariate margins, but its cross-sectional correlation changes sharply from $0.8$ to $-0.8$ when the regime switches from 1 to 2.
Then, by comparing the results of the three bivariate sub-processes, \cref{sec:sim_subset} explores the effects of univariate marginal shift and cross-sectional correlation change on the inference of latent regimes.
Other serial and cross-sectional correlations are chosen to be close to the fitted parameters of the macroeconomic indicators data in \cref{sec:empirical}.
Note that in order to compare the results of the three bivariate sub-processes mentioned above, we make the serial correlations of the univariate components 1 and 3 to be the same as those of the univariate components 2 and 4, respectively.

When the models are fitted, the
skew-t distribution \citep{Jones2003skew_t} is also employed.
The results are based on 100 simulations. In each simulation, new observations are generated and the models are fitted based on the new observations.

\subsection{Fitting with external information on latent regimes}
\label{sec:sim_fit_methods}

In this section,
the estimated parameters based on fitted models with different Markov orders are compared.

To evaluate the fitting of the parameters with external information, we assume the actual latent regime sequence is known, and all parameters are estimated based on the procedure in \cref{sec:estimate_with_external}.
Two models with different Markov order $k_{i,g}=1,2$ for $i=1,\dots,4$
and $g=1,2$ are fitted.
The actual parameters and the means and standard deviations over estimates from 100 simulations are presented in Tables \ref{tb:true_and_estimated_within_regime} and \ref{tb:true_and_estimated_regime_switching}.

\begin{table}[H]
  \small
  \centering
  \begin{tabular}{cccccccc}
  \toprule
  \multirow{3}{*}{\begin{minipage}{0.7in}Parameter notations\end{minipage}} &
  \multicolumn{3}{c}{Regime 1 ($g=1$)} & &
  \multicolumn{3}{c}{Regime 2 ($g=2$)} \\\cmidrule{2-4}\cmidrule{6-8}
   & \multirow{2}{*}{\begin{minipage}{0.5in}True \\values\end{minipage}} & \multicolumn{2}{c}{Estimates} & & \multirow{2}{*}{\begin{minipage}{0.5in}True \\values\end{minipage}} & \multicolumn{2}{c}{Estimates} \\\cmidrule{3-4}\cmidrule{7-8}
   &  & $k_{i,g}=1$ & $k_{i,g}=2$ & &  & $k_{i,g}=1$ & $k_{i,g}=2$  \\\midrule
  \multirow{2}{*}{$\alpha_{1,1,g}$} & \multirow{2}{*}{$0.3$} & $0.30$   & $0.30$    & & \multirow{2}{*}{$0.1$} & $0.09$   & $0.09$ \\
                                    &                        & $(0.05)$ & $(0.05)$  & &                        & $(0.04)$ & $(0.06)$ \\\cdashline{1-8}
  \multirow{2}{*}{$\alpha_{1,2,g}$} & \multirow{2}{*}{$0$}   & ---      & $-0.01$   & & \multirow{2}{*}{$0$}   & ---      & $0.00$ \\
                                    &                        & ---      & $(0.04)$  & &                        & ---      & $(0.04)$ \\\midrule
  \multirow{2}{*}{$\alpha_{2,1,g}$} & \multirow{2}{*}{$0.3$} & $0.29$   & $0.29$    & & \multirow{2}{*}{$0.1$} & $0.10$   & $0.10$ \\
                                    &                        & $(0.06)$ & $(0.06)$  & &                        & $(0.03)$ & $(0.07)$ \\\cdashline{1-8}
  \multirow{2}{*}{$\alpha_{2,2,g}$} & \multirow{2}{*}{$0$}   & ---      & $-0.01$   & & \multirow{2}{*}{$0$}   & ---      & $0.00$ \\
                                    &                        & ---      & $(0.03)$  & &                        & ---      & $(0.04)$ \\\midrule
  \multirow{2}{*}{$\alpha_{3,1,g}$} & \multirow{2}{*}{$0.5$} & $0.49$   & $0.49$    & & \multirow{2}{*}{$0.1$} & $0.09$   & $0.09$ \\
                                    &                        & $(0.05)$ & $(0.05)$  & &                        & $(0.04)$ & $(0.06)$ \\\cdashline{1-8}
  \multirow{2}{*}{$\alpha_{3,2,g}$} & \multirow{2}{*}{$0$}   & ---      & $0.00$    & & \multirow{2}{*}{$0$}   & ---      & $0.00$ \\
                                    &                        & ---      & $(0.04)$  & &                        & ---      & $(0.04)$ \\\midrule
  \multirow{2}{*}{$\alpha_{4,1,g}$} & \multirow{2}{*}{$0.5$} & $0.49$   & $0.49$    & & \multirow{2}{*}{$0.1$} & $0.09$   & $0.09$ \\
                                    &                        & $(0.04)$ & $(0.04)$  & &                        & $(0.04)$ & $(0.06)$ \\\cdashline{1-8}
  \multirow{2}{*}{$\alpha_{4,2,g}$} & \multirow{2}{*}{$0$}   & ---      & $0.00$    & & \multirow{2}{*}{$0$}   & ---      & $0.01$ \\
                                    &                        & ---      & $(0.04)$  & &                        & ---      & $(0.04)$ \\\midrule
  \multirow{2}{*}{$\rho_{1,2,g}$}   & \multirow{2}{*}{$0.3$} & $0.30$   & $0.30$    & & \multirow{2}{*}{$0.1$} & $0.10$   & $0.10$ \\
                                    &                        & $(0.05)$ & $(0.05)$  & &                        & $(0.04)$ & $(0.04)$ \\\cdashline{1-8}
  \multirow{2}{*}{$\rho_{1,3,g}$}   & \multirow{2}{*}{$0.2$} & $0.19$   & $0.19$    & & \multirow{2}{*}{$0.4$} & $0.40$   & $0.39$ \\
                                    &                        & $(0.06)$ & $(0.06)$  & &                        & $(0.03)$ & $(0.03)$ \\\cdashline{1-8}
  \multirow{2}{*}{$\rho_{1,4,g}$}   & \multirow{2}{*}{$0.2$} & $0.18$   & $0.18$    & & \multirow{2}{*}{$0.1$} & $0.10$   & $0.10$ \\
                                    &                        & $(0.06)$ & $(0.06)$  & &                        & $(0.04)$ & $(0.04)$ \\\cdashline{1-8}
  \multirow{2}{*}{$\rho_{2,3,g}$}   & \multirow{2}{*}{$0.3$} & $0.30$   & $0.30$    & & \multirow{2}{*}{$0.2$} & $0.20$   & $0.20$ \\
                                    &                        & $(0.06)$ & $(0.06)$  & &                        & $(0.04)$ & $(0.04)$ \\\cdashline{1-8}
  \multirow{2}{*}{$\rho_{2,4,g}$}   & \multirow{2}{*}{$0.2$} & $0.20$   & $0.20$    & & \multirow{2}{*}{$0.1$} & $0.10$   & $0.10$ \\
                                    &                        & $(0.06)$ & $(0.06)$  & &                        & $(0.04)$ & $(0.04)$ \\\cdashline{1-8}
  \multirow{2}{*}{$\rho_{3,4,g}$}   & \multirow{2}{*}{$0.8$} & $0.79$   & $0.79$    & & \multirow{2}{*}{$-0.8$}& $-0.80$  & $-0.80$ \\
                                    &                        & $(0.02)$ & $(0.02)$  & &                        & $(0.01)$ & $(0.01)$ \\
  \bottomrule
  \end{tabular}
  \captionsetup{font=footnotesize}
  \caption{True values and estimates of partial serial autocorrelations and cross-sectional correlations of 4 univariate components in two regimes, given the true latent regime sequence.
            For each parameter, the presented values are mean and standard deviation (in brackets) over parameters estimates from 100 simulations, in each of which two models assuming Markov order $k_g=k_{i,g}=1,2$ for $i=1,\dots,4$ and $g=1,2$ are fitted using the procedure in \cref{sec:estimate_with_external}. From Markov chain theory, the expected number of regime switches and expected time in regime 1 are 29 and 286, respectively.
}
  \label{tb:true_and_estimated_within_regime}
\end{table}

As described in \cref{sec:estimate_with_external}, after estimating the marginal parameters, we take the advantage of closure under margins for a given regime by first estimating the within-regime serial correlations of each univariate component separately, which is followed by estimating the cross-sectional correlation within a regime.
\cref{tb:true_and_estimated_within_regime} presents the results of partial autocorrelations and cross-sectional correlations for each regime.
The estimates are close to the actual values for both partial autocorrelations and cross-sectional correlations.
Moreover, the estimates of partial autocorrelations of lags larger than $1$ are close to $0$ even though all univariate components are fitted separately.
This is because the model used to simulate the data has univariate
margins that are Markov of order 1.
The results demonstrate good performance of the multi-stage fitting procedure.

For the parameters handling transitions between regimes, \cref{tb:true_and_estimated_regime_switching} gives the results for different Markov orders $k_{i,g}=1,2$.
One can see that the mean estimates are also close to the actual values.
To further investigate the effect of fitting the regime-switching serial correlations, the AIC values of models considering and not considering the regime-switching serial correlations are computed and compared.
The model not considering the serial correlations during regime switching corresponds to the special case with $\rho_1=\dots=\rho_4=0$, i.e., the observations after regime switching are independent of those before.
Thus, the model that does not consider regime-switching serial correlations has $4$ fewer parameters than the general model in \cref{def:model}.
The average values of AIC are presented in \cref{tb:AIC_with_and_with_out_rho}.
Note that the AIC values  are calculated based on the complete likelihood of the observations (with the true latent regime sequence assumed known).
The average values of AIC in \cref{tb:AIC_with_and_with_out_rho} indicate that
introducing the parameters for regime-switching serial correlations can lead
to a better fitting model as it results in smaller AIC values in all cases of $k_{i,g}=1,2$ for $i=1,\dots,4$ and $g=1,2$.

\begin{table}[H]
  \small
  \centering
  \begin{tabular}{cccc}
  \toprule
  \multirow{2}{*}{Parameter notations}     & \multirow{2}{*}{True values} & \multicolumn{2}{c}{Estimates} \\\cmidrule{3-4}
                              &                        & $k_{i,g}=1$ & $k_{i,g}=2$  \\\midrule
  \multirow{2}{*}{$\rho_{1}$} & \multirow{2}{*}{$0.1$} & $0.10$      & $0.09$  \\
                              &                        & $(0.07)$    & $(0.07)$  \\\cdashline{1-4}
  \multirow{2}{*}{$\rho_{2}$} & \multirow{2}{*}{$0.2$} & $0.19$      & $0.18$  \\
                              &                        & $(0.11)$    & $(0.12)$  \\\cdashline{1-4}
  \multirow{2}{*}{$\rho_{3}$} & \multirow{2}{*}{$0.1$} & $0.10$      & $0.11$  \\
                              &                        & $(0.04)$    & $(0.04)$  \\\cdashline{1-4}
  \multirow{2}{*}{$\rho_{4}$} & \multirow{2}{*}{$0.2$} & $0.19$      & $0.19$  \\
                              &                        & $(0.04)$    & $(0.04)$  \\
  \bottomrule
  \end{tabular}
  \captionsetup{font=footnotesize}
  \caption{True values and estimates of serial correlations of 4 univariate components during regime switching, given the true latent regime sequence.
For each parameter, the presented values are mean and standard deviation (in brackets) over parameter estimates from 100 simulations, in each of which two models of different Markov order $k_{i,g}=1,2$ for $i=1,\dots,4$ and $g=1,2$ are fitted based on the procedure in \cref{sec:estimate_with_external}.}
  \label{tb:true_and_estimated_regime_switching}
\end{table}

\begin{table}[H]
  \small
  \centering
  \begin{tabular}{ccc}
  \toprule
  & $k_{i,g}=1$ & $k_{i,g}=2$  \\\midrule
  Models without serial correlation during regime switching &
 $13.74$ & $13.77$ \\
  Models with serial correlation during regime switching &
 $13.71$ & $13.73$ \\
  \bottomrule
  \end{tabular}
  \captionsetup{font=footnotesize}
  \caption{Average values of AIC/$T$=AIC/1000 over $100$ simulations, for models with and without serial correlation during regime switching, given the true latent regime sequence.
 In each simulation, two models of different Markov order $k_{i,g}=1,2$ for $i=1,\dots,4$ and $g=1,2$ are fitted according to \cref{sec:estimate_with_external}.
The values are computed based on the complete likelihood of the known latent regime sequence and observations.}
  \label{tb:AIC_with_and_with_out_rho}
\end{table}

\subsection{Study on variable subset selection}
\label{sec:sim_subset}

This section explores the effects of univariate component subset selection by fitting the models using method in \cref{sec:multi-stage_approach} with different bivariate sub-processes and comparing their results.

As mentioned in the previous section, a strategy for reducing the computational cost in cases of high-dimensional time series is to infer the latent regime sequence by fitting a model on a subset of selected univariate time series components.
Then, the parameters of the remaining univariate components can be estimated by treating the inferred latent regime sequence as external information or a proxy of the unknown actual regime sequence.
In this part, the feasibility of this idea is explored with one simulated series by investigating the effects of different univariate component subsets on inference of the latent regime sequence.

\begin{figure}[H]
  \centering
  \includegraphics[scale=0.5]{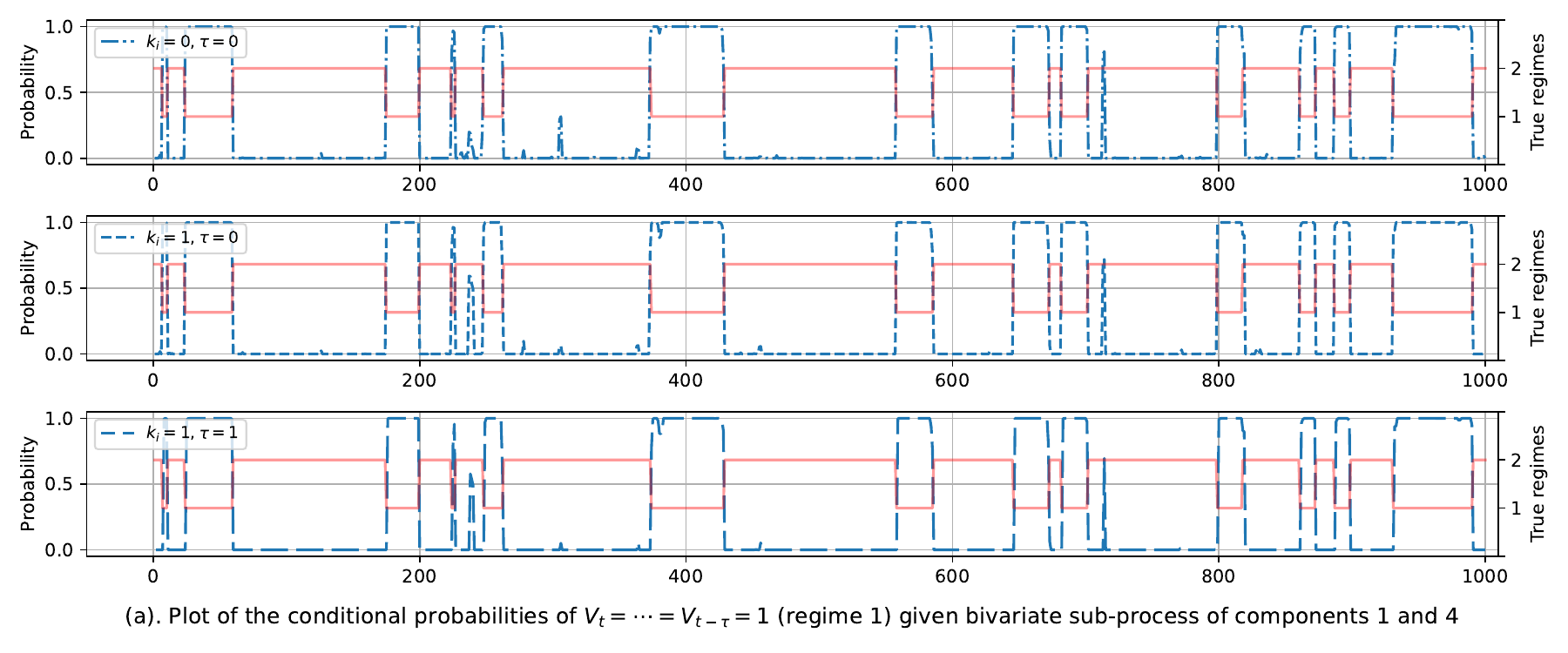}
  \includegraphics[scale=0.5]{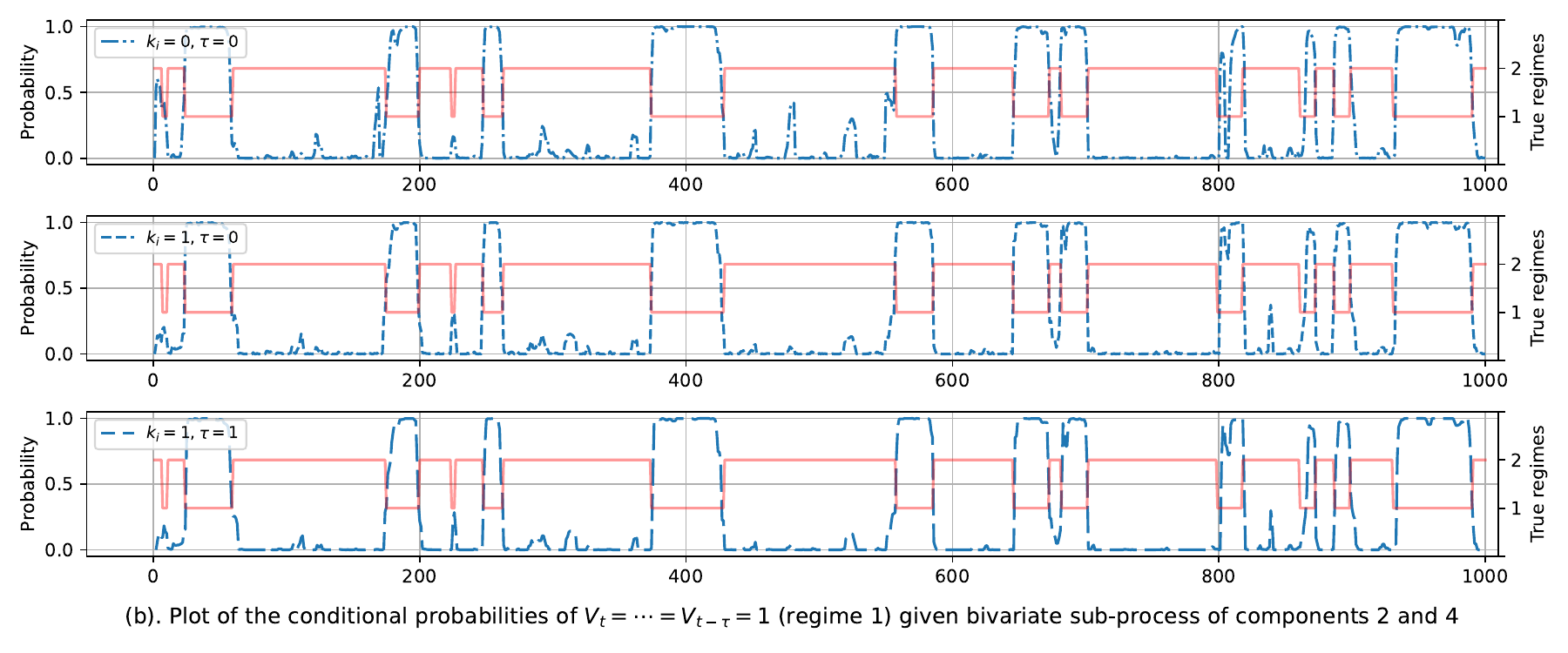}
  \includegraphics[scale=0.5]{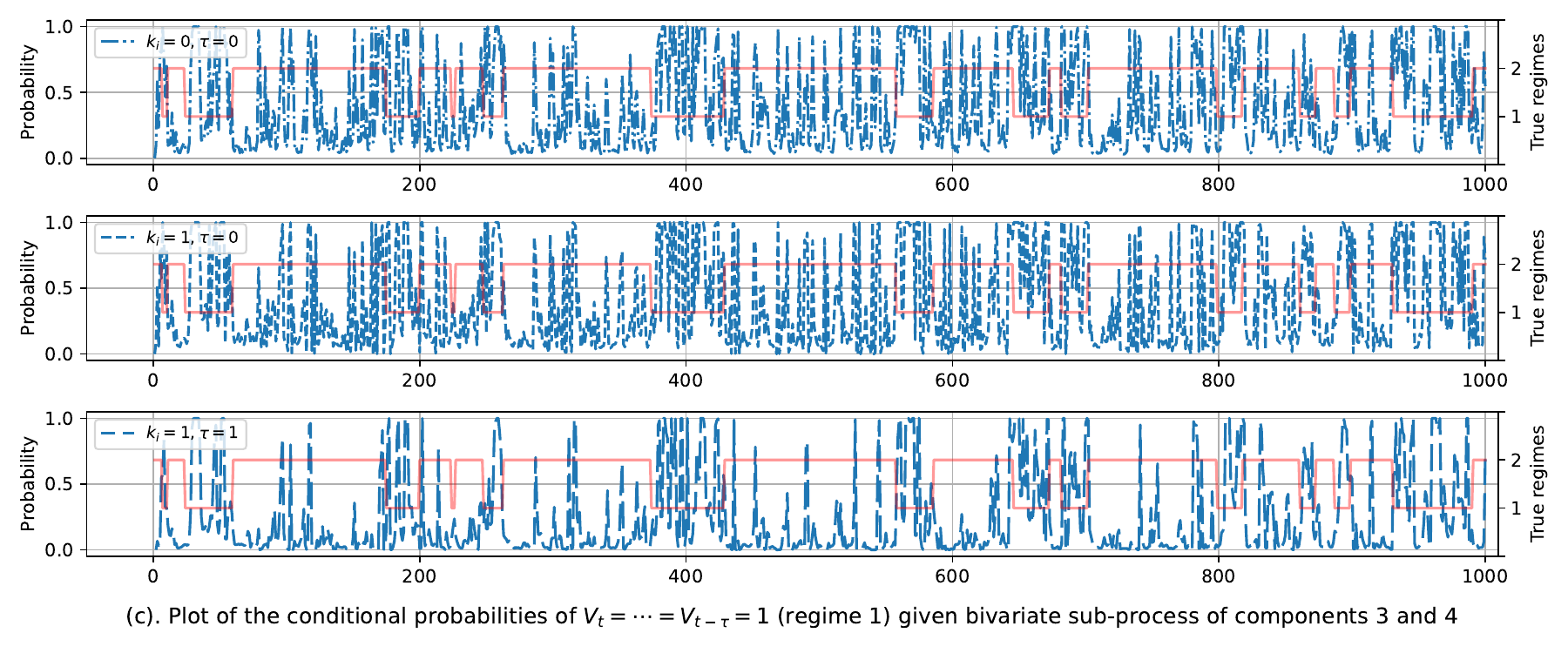}
  \captionsetup{font=footnotesize}
  \caption{The conditional probabilities of $V_t=\cdots=V_{t-\tau}=1$ (regime 1) given different bivariate sub-processes, with $k_g=k_{i,g}=0, \tau=0$; $k_g=k_{i,g}=1,\tau=0$; and $k_g=k_{i,g}=1,\tau=1$ for $i=1,\dots,4$ and $g=1,2$.
The conditional probabilities can be derived by fitting a model with the given bivariate sub-process and computing the conditional probabilities of regimes given the bivariate sub-process based on the fitted model.
The solid line indicates the true latent regime sequence.}
  \label{fig:inferred_probs}
\end{figure}

With the simulated 4-dimensional series introduced in \cref{sec:sim_fit_methods}, three bivariate sub-processes are considered: the subprocess composed of univariate components 1 and 4, the subprocess composed of components 2 and 4, and the subprocess composed of components 3 and 4.
The choice of these three sub-processes is made in order to analyze influences of shifts in  univariate margins and changes in cross-sectional correlations on the detection of latent regime switches. According to \cref{tb:true_and_estimated_within_regime}, when switching from regime 1 to regime 2, the location parameters of margins shift significantly for univariate component 1, and shift only slightly for univariate component 2.
While the cross-sectional correlations between components 1 and 4, and between components 2 and 4 only change a little from $0.3$ to $0.1$ when the regime switches, the cross-sectional correlations between components 3 and 4 change sharply from $0.8$ to $-0.8$.
The reason why we focus on the univariate margins is that their estimates
of $\{k_{i,g}\},\btheta_{\boldeta},\btheta_V$
can be obtained by fitting univariate hidden Markov models.
The estimates from the univariate hidden Markov models can be used as a reference when the selection of univariate components is performed.
For each bivariate sub-process, two models with $k_g=k_{i,g}=0$ and $k_g=k_{i,g}=1$ for $1\leq i\leq d$ and $1\leq g\leq G$ are fitted so that the effects of fitting the serial correlations can also be roughly explored.
Note that in the case when $k_1=k_2=0$, we also let $P=\bm{0}$ in order to remove all serial correlations and the model in this case is indeed equivalent to a bivariate hidden Markov model.
To visualize the ability of inferring the latent regime sequence, the conditional probabilities of $V_t=\cdots=V_{t-\tau}=1$ given the fitted two univariate components for $t=\tau+1,\dots,1000$ are derived for each model.
In \cref{fig:inferred_probs}, the sequence of the derived conditional probabilities in the cases of $k_g=k_{i,g}=0, \tau=0$; $k_g=k_{i,g}=1, \tau=0$; and $k_g=k_{i,g}=1, \tau=1$ for $i=1,\dots,4$ are shown for each bivariate sub-process (dashed lines), along with the actual regime sequence (solid line).
\cref{fig:inferred_probs} (c) shows that for the sub-process composed of components 3 and 4, the latent regime switching cannot be correctly detected using the conditional probabilities.
However, the results of latent regimes detection are better for the sub-process composed of components 2 and 4, and the latent regimes can be accurately inferred when it comes to the sub-process composed of components 1 and 4.
It indicates that a model based on a sub-process whose univariate margins are similar across different regimes fails to detect the latent regime switching, even though there is a large difference in its cross-sectional correlations in different regimes.
By comparing three subplots in \cref{fig:inferred_probs} (a), one can see that fitting the serial correlation can slightly improve the results as the observations are simulated from a model of Markov order 1 within each regime.
Moreover, the third subplot for $\tau=1$ shows a much smoother probability curve.
\begin{figure}[h]
  \centering
  \includegraphics[scale=0.45]{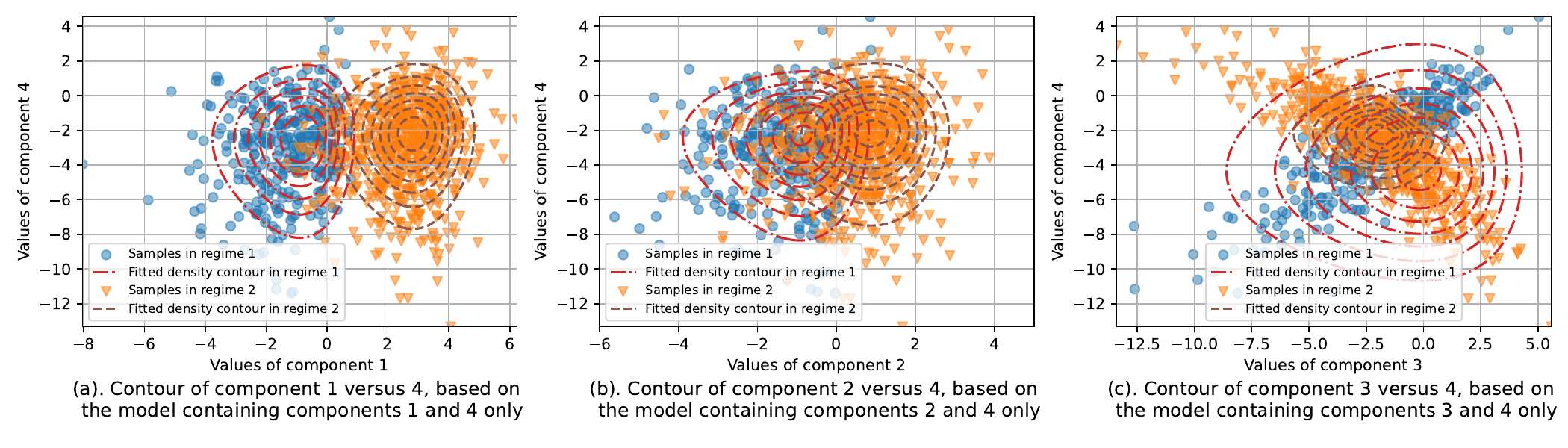}
  \includegraphics[scale=0.45]{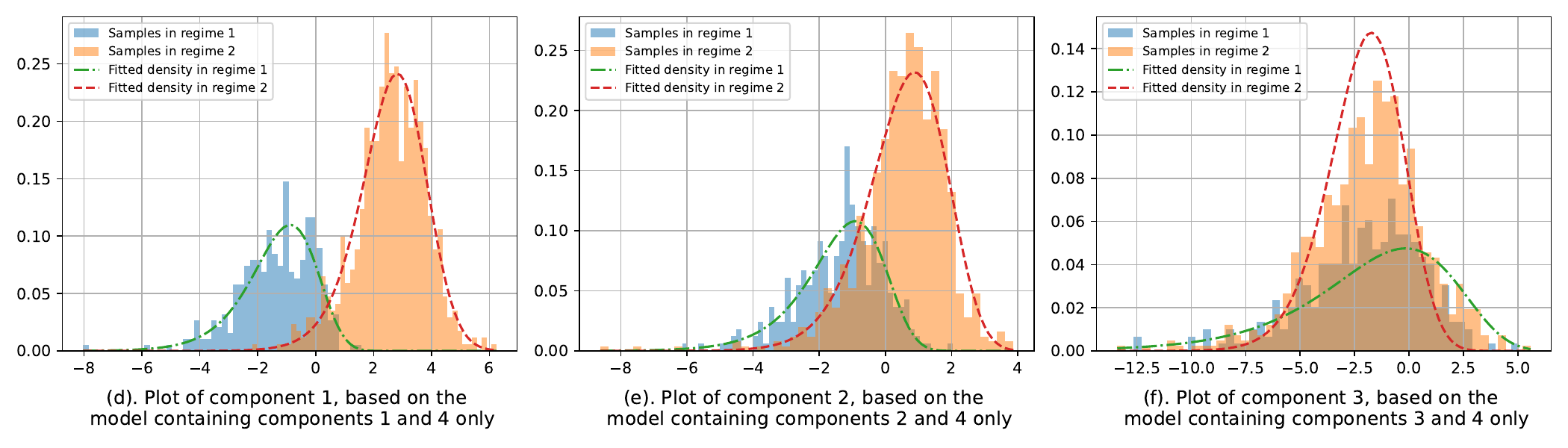}
  \captionsetup{font=footnotesize}
  \caption{Plots (a) to (c): Scatterplots of observations and contour plots of fitted joint densities of different bivariate sub-processes.
The fitted distributions are from the models fitting three subprocesses with Markov order $k_{i,g}=1$ for all $i=1,\dots,4$ and $g=1,2$.
Plots (d) to (f): Univariate marginal histograms and fitted densities of components 1-3.}
  \label{fig:inferred_margins}
\end{figure}

Using \cref{fig:inferred_margins}, a general conclusion can be drawn that the greater the distance between the univariate margins in different regimes, the better the inference for the latent regime sequence based on the sub-process involving these marginal components.
The marginal density and contour plots in \cref{fig:inferred_margins} can provide an intuitive explanation.
Subplots (a) to (c) show the scatterplots and contours of the fitted densities for the considered three bivariate sub-processes, while subplots (d) to (f) display the univariate marginal histograms and fitted densities of univariate components 1 to 3.
From subplots (a) and (d), one can observe that, for the bivariate sub-process composed of components 1 and 4, the observations in different regimes are well-separated. This is due to a large difference in the location parameters of the underlying distribution for marginal component~1 under the two regimes. As a result, the bivariate joint distributions of this sub-process can be accurately fitted in both regimes, even though there is no shift in margins under the two regimes for component 4. This also explains why the latent regime sequence can be correctly inferred using the sub-process with these two marginal components. For the bivariate sub-process composed of components 2 and 4, even though there is some distance between location parameters of the univariate margins of component 2 and the model can separate observations in the different regimes to some extent, the accuracy in inferring the latent regime sequence is diminished. Finally, for the last bivariate sub-process, plot~(c) reveals that the model fails to identify the distributions in the two regimes even though their correlation parameters are quite different. This can be attributed to the lack of differences in the marginal distributions across the regimes as shown in plot~(f). This failure of the model to discriminate between distributions for components 3 and 4 in the two regimes leads to the model's inability to accurately infer the latent regime sequence when components 3 and 4 are used.

This study suggests that the subset selection of univariate components for the initial inference of the latent regime sequence should be based on those components whose univariate margins in different regimes can be well discriminated by a univariate hidden Markov model.


\section{An empirical study on macroeconomic business cycle}
\label{sec:empirical}

This section contains the application of the margin-closed regime-switching model to the time series of macroeconomic indicators from the FRED-MD database \citep{Mccracken2016fred}.
The database contains monthly series of many macroeconomic variables. \cref{sec:data} briefly introduces the data set.
\cref{sec:empirical_with_NBER} fits the model with the given external business cycles  information.
\cref{sec:empirical_without_NBER} identifies the business cycle based only on the observed time series.

\subsection{Macroeconomic indicators and business cycle}
\label{sec:data}

An available source of the business cycle information is the National Bureau of Economic Research (NBER, \url{https://www.nber.org/research/data/us-business-cycle-expansions-and-contractions}) of the United States.
It is a nonprofit research organization serving a beneficial role in cataloging stylized facts about business cycles and providing a historical accounting of the dates of regime shifting for economic growth.
For the macroeconomic indicators, the primary assumption that the series is stationary within each regime may be reasonable since all variables in the database have been transformed through differencing of natural log.
To infer latent business cycle by fitting the margin-closed model, a subset of the indicators in the FRED database should be selected.
Here, we choose the four key indicators mentioned by \cite{chauvet2005}: total personal income less transfer payments (\texttt{income}), the growth rates of manufacturing and trade sales (\texttt{sales}), civilian labor force employed in nonagricultural industries (\texttt{employment}), and industrial production (\texttt{IP}).
All these variables are transformed to the first difference of natural log. The time series are available from June 1961 to February 2020, inclusive, for a total of 705 months.

\subsection{Inference with external information}
\label{sec:empirical_with_NBER}

We first fit the proposed margin-closed regime-switching model to the selected four macroeconomic indicators and using the external business cycle information released by the NBER, with two regimes for economic recession and expansion.


As the univariate margins may have heavier tails and skewness compared with the Gaussian distribution, we fit them with the
skew-t distribution \citep{Jones2003skew_t} and transform to have the standard normal distribution.

For simplicity, the within-regime Markov orders for all univariate components are set to be equal, i.e., $k_{i,g}=k_g=k-1$ for all $i=1,\dots,4$ and $g=1,2$. In this case, one can take $k$ as $1$ plus the largest Markov order over all univariate time series.
To determine the Markov order $k$ of the model, using the business cycle information from the NBER as a proxy for the latent regime sequence,  we examined the plots of the sample autocorrelation function (ACF) and partial autocorrelation function (PACF). These plots (not shown in the paper) suggest that the largest Markov order across all univariate time series and all regimes is $4$.
The AIC values of the models with different Markov orders are given in \cref{tb:AIC_with_known_latents}.
Note that $k_{i,g}=0$ for $i=1,\dots,4$ and $g=1,2$ in the table indicates the hidden Markov model, where we also set $P=\bm{0}$.
The results in \cref{tb:AIC_with_known_latents} show that the models with $k_{i,g}=3,4,5$ have much smaller AIC values than those with $k_{i,g}=0,1,2$.
This is consistent with the earlier assertion that $k_{i,g}=4$ based on the sample ACF and PACF plots.
In the subsequent analysis, we set $k_{i,g}=3$ and $k=4$ for $i=1,\dots,4$ and $g=1,2$ as an optimal choice leading to the smallest AIC value.

\begin{table}[H]
  \small
  \centering
  \begin{tabular}{ccccccc}
    \toprule
     Within-regime Markov order $k_{i,g}$ ($i=1,\dots,4,g=1,2$) &       0 &       1 &       2 &       3 &       4 &      5 \\
    \midrule
     AIC & 3884 & 3804 & 3789 & 3768 & 3771 & 3776 \\
    \bottomrule
    \end{tabular}
  \captionsetup{font=footnotesize}
  \caption{The AIC values of models fitted with business cycle information given by NBER. The values are computed based on the complete likelihood of the known latent regime sequence and observations.
Markov order of 0 leads to the hidden Markov model.}
  \label{tb:AIC_with_known_latents}
\end{table}

\begin{figure}[H]
  \centering
  \includegraphics[scale=0.45]{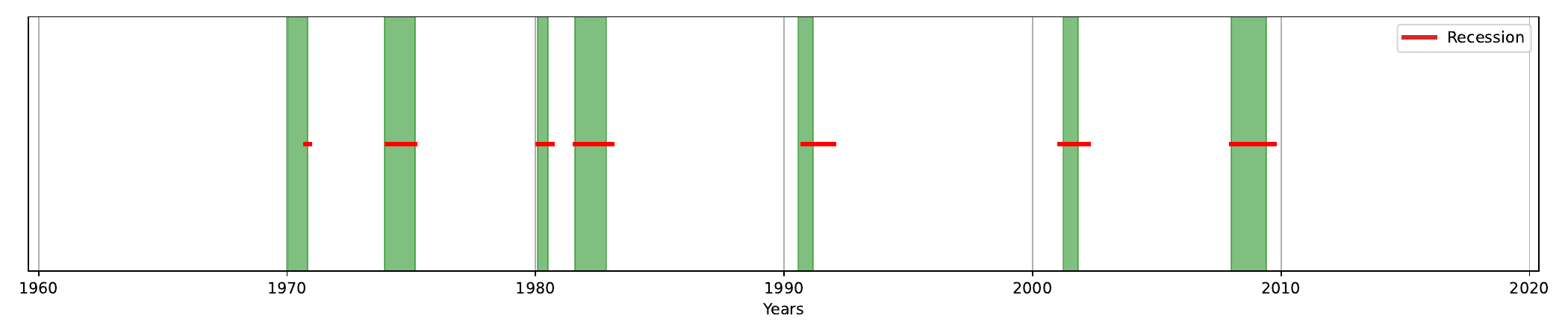}
  \captionsetup{font=footnotesize}
  \caption{Inferred business cycles from the model with $\tau=0$, $\nu=3$, and $\xi=0.8$, with the external business cycle information given by NBER.
The red solid lines denote the inferred recession period, and the green shaded areas denote the NBER recession dates.}
  \label{fig:inferred_regimes_with_known_latents}
\end{figure}

After determining the Markov order, we update the latent regime sequence (business cycle) following the approach in \cref{sec:latent_infer_with_external}.
It runs iteratively until the regime sequence stabilizes. For smoothing parameters in the latent regime sequence updating, we chose the same values as in \citet{chauvet2008comparison}, i.e., $\tau=0$, $\nu=3$, and $\xi=0.8$.
The results of the inferred business cycle are presented in \cref{fig:inferred_regimes_with_known_latents}.
From the plot, we can see that the inferred business cycle updated based on the model has the same number of recession periods as that from the NBER. The differences include a delayed start and shorter duration of the first recession period, but longer recessions after the second recession period.

%

\subsection{Inference based on the observed macroeconomic indicators only}\
\label{sec:empirical_without_NBER}

We now fit the margin-closed regime-switching model and infer the business cycle based only on the observations of the four macroeconomic indicators.

We first fit the model using the multi-stage approach described in \cref{sec:multi-stage_approach}.
In this case, both the Markov order $k$ and the number of different regimes $G$ should be determined for the model.
Since the business cycle is usually modelled by two states of expansion and recession, we consider a model with three regimes for comparison.
For all models, we let $g=1$ represent the recession and let higher values of $g$ indicate better economic situations.
Therefore, models with two and three regimes and within-regime Markov orders from $0$ to $5$ are considered.
The AIC values for the fitted models are summarized in \cref{tb:AIC_without_latents}.
Note that when $k_{i,g}=0$ for all $1\leq i\leq 4$ and $1\leq g\leq 3$, we also let $P=\bm{0}$ so that the model reduces to a hidden Markov model.
According to the AIC values, models having the within-regime Markov order of $3$ are preferred in both cases of 2 and 3 regimes.

\begin{table}[H]
  \small
  \centering
  \begin{tabular}{cccccc}
    \toprule
     Within-regime Markov order $k_{i,g}$ ($i=1,\dots,4,g=1,2,3$)   &       0 &       1 &       2 &       3 &       4 \\
    \midrule
     AIC of models with two regimes      & 3897 & 3789  & 3768 & 3753 & 3759  \\
     AIC of models with three regimes      & 3812 & 3692  & 3679 & 3676 & 3689 \\
    \bottomrule
    \end{tabular}
  \captionsetup{font=footnotesize}
  \caption{The AIC values of models fitted by multi-stage approach, without external business cycle information. Markov order of 0 in the table corresponds to the hidden Markov model.}
  \label{tb:AIC_without_latents}
\end{table}

\begin{figure}[H]
  \centering
  \includegraphics[scale=0.5]{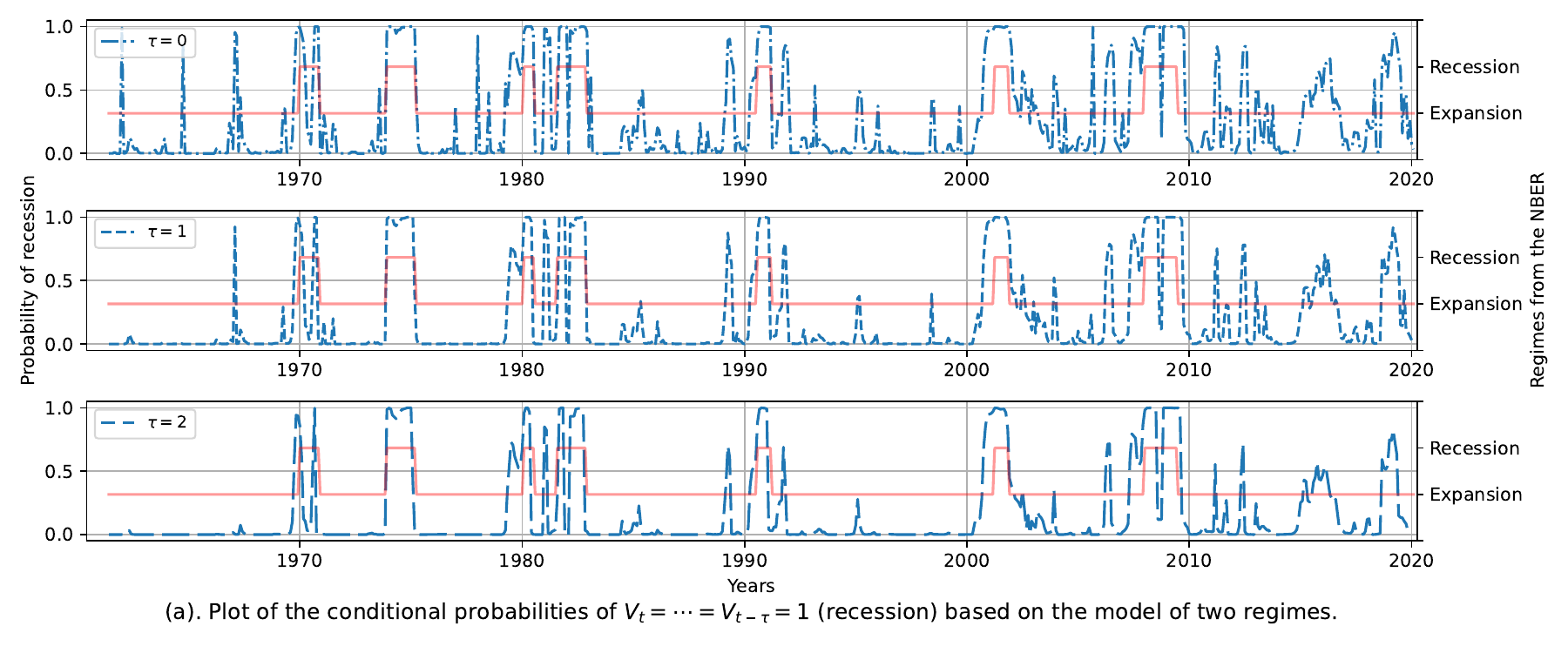}
  \includegraphics[scale=0.5]{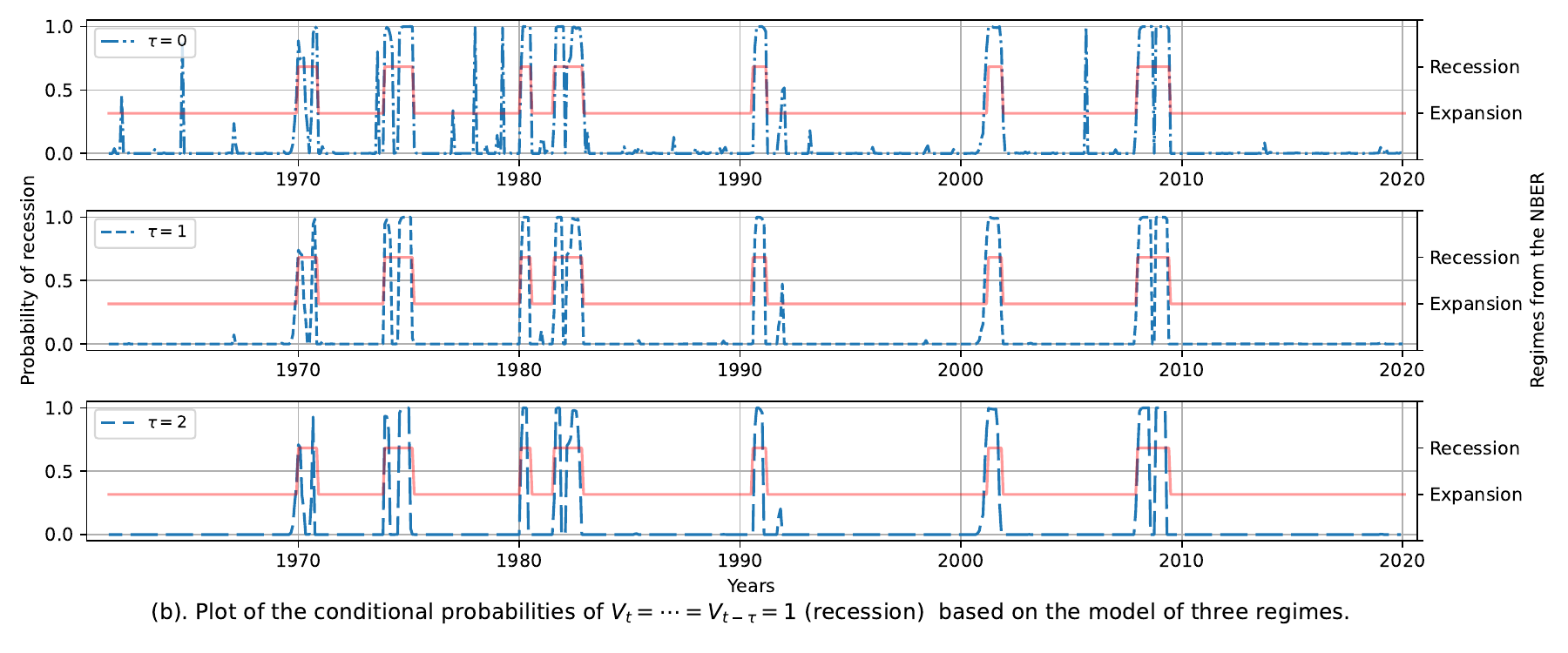}
  \captionsetup{font=footnotesize}
  \caption{Inferred conditional probability of recession given all observations for the four indicators, based on models with two regimes (panel~(a)) and three regimes (panel~(b)).}
  \label{fig:unknown_latents_probs}
\end{figure}

To further compare the two models with two and three regimes, \cref{fig:unknown_latents_probs} shows the estimates of the conditional probability of business recession given all observations, i.e., function $p_{t,\tau}(g)$ with $g=1$ and $\tau=0,1,2$ in the considered models.
If we use the business cycles  from the NBER as the benchmark, the plots show closer matches for the model with three regimes than the model with two regimes.
Moreover, for both models, increasing the smoothing parameter $\tau$ from
$0$ to $1$ can improve the results as the fluctuations of the curves is reduced.
But the benefit from further increasing $\tau$ to $2$ is minor as the curves corresponding to $\tau=1$ are already sufficiently smooth.

We then apply the iterative procedure described in \cref{sec:iterative_approach}.
Specifically, we employ the results of the hidden Markov model with independent univariate random variables in emission distributions as the starting point.
Then, given the hyper-parameters $\tau,\nu,\xi$ in latent business cycle inference, the estimates of all model parameters and the inferred business cycles  are updated iteratively until they stabilize.
According to the outcomes shown in \cref{fig:unknown_latents_probs}, $\tau=1$ is an appropriate choice for both models with two and three regimes.

\begin{figure}[H]
  \centering
  \includegraphics[scale=0.45]{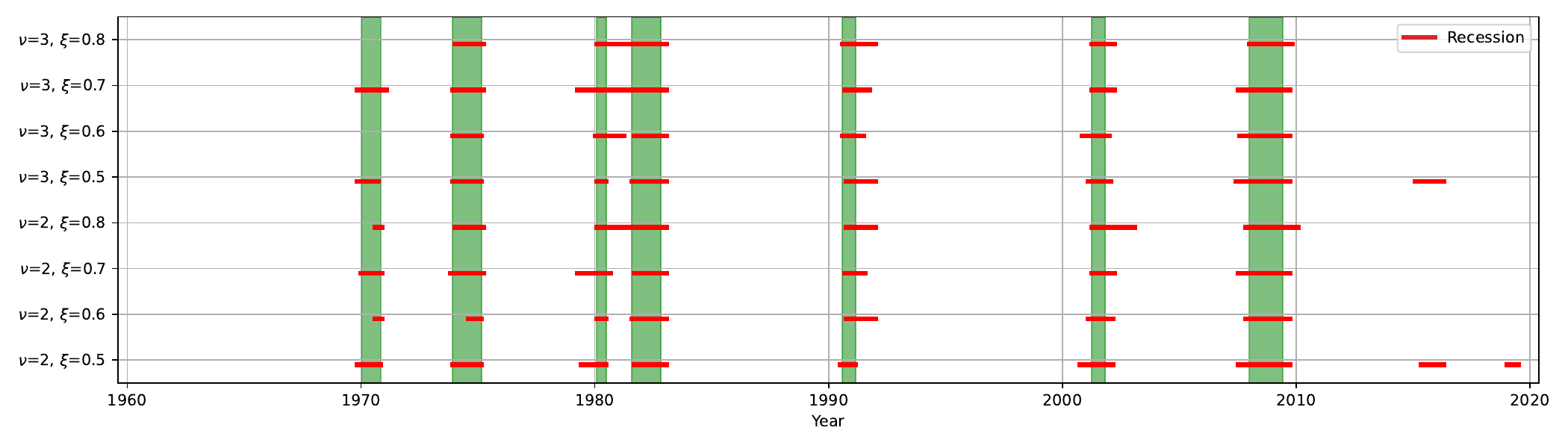}
  \captionsetup{font=footnotesize}
  \caption{Inferred latent regimes for parameter $\tau=1$ and different values $\nu, \xi$, based on the models with two regimes fitted using observations on four macroeconomic indicators.
The red solid lines denote the inferred period of recession and the green shaded areas denote the NBER recession dates.}
  \label{fig:inferred_regimes2}
\end{figure}

\cref{fig:inferred_regimes2} presents the results of the model with two regimes.
One can see that the inferred business cycles are sensitive to the parameters $\nu$ and $\xi$. Compared with the business cycles  given by the NBER, the large $\xi$ value of $0.8$ leads to a somewhat inadequate result as the fitted model fails to separate the third and fourth recession periods.
Also, for $\nu=3$, the fitted model tends not to identify the first recession period. These results suggest that the value $\nu=2$ and a smaller $\xi$ value are preferable.

\begin{figure}[H]
  \centering
  \includegraphics[scale=0.45]{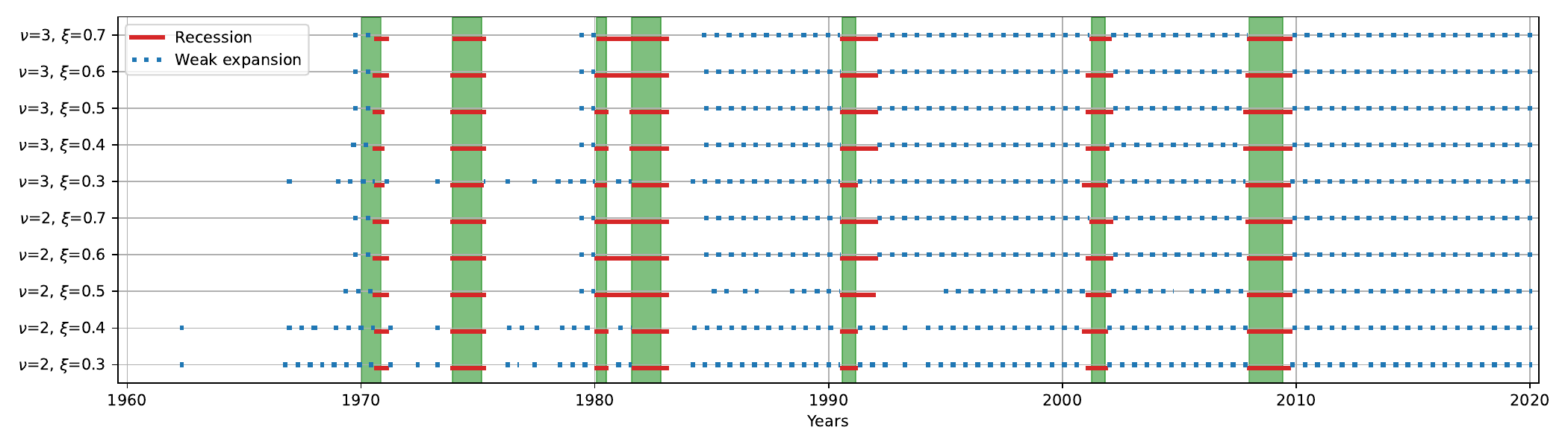}
  \captionsetup{font=footnotesize}
  \caption{Inferred latent regimes for parameter $\tau=1$ and different values of $\nu, \xi$, based on the models with three regimes (recession, weak expansion, strong expansion) fitted using only observations on four macroeconomic indicators.
The red solid lines denote the inferred periods of recession, the blue dot lines denote the inferred periods of weak expansion, and the green shaded areas denote the NBER recession dates.}
  \label{fig:inferred_regimes3}
\end{figure}

For the models with three regimes, we interpret the three states within a business cycle as recession, weak expansion, and strong expansion.
\cref{fig:inferred_regimes3} displays the inferred latent regimes from the model with three regimes for different values of $\nu$ and $\xi$.
Note that the red solid lines denote the inferred periods of recession, the blue dot lines indicate periods of weak expansion, and blank periods are of strong expansion.
Since there are three regimes in the model, we also try smaller values of $\xi$.
Unlike the results of the model with two regimes, the inferred business cycles  from the models with three regimes are more stable with respect to hyper-parameters $\nu$ and $\xi$.
The main problem is that higher values of $\xi$ tend to lead to longer periods of the fifth recession and fail in separating the third and fourth recession periods.
Moreover, the inferred starting time of the first recession period is still delayed compared with the period given by the NBER.

Based on \cref{fig:inferred_regimes3}, one can notice that the results for $\nu=2$ and $\xi=0.3$ can best match the business cycles  from the NBER.
To better explain this result, the mean and mode values of the four transformed macroeconomic indicators in different regimes are given in \cref{tb:mean_and_mode}.
It can be noticed that the means and modes of all indicators are negative in the recession, and they are all positive in the other two regimes, corresponding to $g=2$ and $g=3$. That is why $g=2$ and $g=3$ are interpreted as expansion, but the expansion indicated by $g=2$ is weaker than that of $g=3$.
To compare the three regimes, \cref{tb:mean_and_mode} shows the cross-sectional correlation matrices of the four macroeconomic indicators in different regimes.
They are the correlation matrices of the $4$-dimensional Gaussian copulas of the transformed indicators' contemporaneous joint distributions in different regimes.
It is seen that generally the four transformed indicators are more correlated in the recession and strong expansion periods than they are in weak expansion,
except for the contemporaneous correlation between sales and employment, and the contemporaneous correlation between income and IP.
It reflects that those macroeconomic indicators are more correlated in extreme economic situations.
The interpretation of the inferred latent regimes  is that even though the economy could be in expansion, most expansion periods after 1967 were of a weak expansion when those indicators had lower increment rates than the strong expansion before 1967.

\begin{table}[H]
  \small
  \centering
  \begin{tabular}{cccccccccccc}
    \toprule
                            & \multicolumn{2}{c}{Income} &&  \multicolumn{2}{c}{Sales} &&  \multicolumn{2}{c}{Employment} && \multicolumn{2}{c}{IP} \\
                            & Mean  &   Mode &&   Mean &   Mode &&   Mean &   Mode &&   Mean &   Mode \\\cmidrule{2-3}\cmidrule{5-6}\cmidrule{8-9}\cmidrule{11-12}
    Recession ($g=1$)       & -0.18 &  -0.11 &&  -0.38 &  -0.56 &&  -0.80 &  -0.54 &&  -0.08 &  -0.05 \\
    Weak expansion ($g=2$)  &  0.29 &   0.29 &&   0.23 &   0.19 &&   0.21 &   0.23 &&   0.14 &   0.15 \\
    Strong expansion ($g=3$)&  0.38 &   0.47 &&   0.71 &   1.04 &&   0.76 &   0.74 &&   0.22 &   0.22 \\
    \bottomrule
    \end{tabular}
  \captionsetup{font=footnotesize}
  \caption{The mean and mode values of the four transformed macroeconomic indicators (\%) in different regimes inferred using the model with three regimes.}
  \label{tb:mean_and_mode}
\end{table}

\begin{table}[H]
  \small
  \centering
  \begin{tabular}{cccc}
  \toprule
            & Recession ($g=1$) & Weak expansion ($g=2$) & Strong expansion ($g=3$) \\\midrule
  $R_{\bY,g}$ &
  $\bsmat
    1 &    0.06 &    0.18 &    0.36\\
    0.06 &    1 &    0.51 &    0.18\\
    0.18 &    0.51 &    1 &    0.21\\
    0.36 &    0.18 &    0.21 &    1
  \esmat$ &
  $\bsmat
    1 &    0.07 &    0.16 &    0.17\\
    0.07 &    1 &    0.40 &    0.08\\
    0.16 &    0.40 &    1 &    0.13\\
    0.17 &    0.08 &    0.13 &    1
  \esmat$ &
  $\bsmat
    1 &    0.28 &    0.26 &    0.13\\
    0.28 &    1 &    0.18 &    0.26\\
    0.26 &    0.18 &    1 &    0.38\\
    0.13 &    0.26 &    0.38 &    1
  \esmat$\\
  \bottomrule
  \end{tabular}
  \captionsetup{font=footnotesize}
  \caption{The contemporaneous cross-sectional correlations of the four macroeconomic indicators in different regimes inferred using the model with three regimes.}
  \label{tb:contemp_corr}
\end{table}

\section{Discussion}
\label{sec:discussion}

The proposed margin-closed regime-switching multivariate time series
model of Markov order $k$ implies that any subprocess of the multivariate
time series is a regime-switching model
with the same Markov order $k$ and the same latent regime sequence.
It leads to a parsimonious regime-switching model and allows to make inference for the latent regime sequence based on a sub-process composed of some selected univariate components. \cref{sec:simulation} suggests selecting components that have the largest differences in univariate distributions under the different regimes.
The proposed multivariate margin-closed regime-switching model is applied to a data set of four macroeconomic indicators, and the results show good performance on the inference for the latent business cycle.

One potential drawback of the proposed model is that the closure under margins property restricts the dependence between observations in different regimes. One way to deal with this is to introduce more parameters to capture serial dependence when transitioning between regimes.
Alternatively, one can remove the restriction of closure under margins. This may be suitable when the dimension of the time series is low and the series is sufficiently long. A more general dependence between observations in different regimes can then be constructed through conditional independence, i.e., Condition 3 in \cref{def:model} can be replaced with the assumption that
\[
\sbra{\bY_{t^*+\tilde{\ell}} \bot \bY_{t^*-\ell}}\big|\bY_{t^*+\tilde{\ell}-1},\dots,\bY_{t^*-\ell+1},\quad \forall\ \ell,\tilde{\ell}\geq 0\ \mbox{and}\ \tilde{\ell}+\ell>k, \ V_{t^*}\ne V_{t^*-1}.
\]
where $k$ is the Markov order of $\cbra{\bY_{t}}_{t\in\ints}$, and only the correlation matrices
$\cov\rbra{\bY_{t^*+\tilde{\ell}},\bY_{t^*-\ell}}$ for $\ell, \tilde{\ell}\geq 0$ and $\tilde{\ell}+\ell<=k$ need to be parameterized. In this case the dependence between observations in different regimes is fully modelled. Some advantages of this model formulation compared with other regime-switching models such as Markov switching vector autoregressive (MSVAR) models (see  \cite{Cheng2016}, \cite{Sola1994}, and \cite{Hamilton1990}) include the availability of thr stationary joint distributions of observations within each regime and the possibility to be extended to models with non-Gaussian margins.

%

\begin{appendices}

\section{Parameterization of margin-closed VAR model}
\label{ap: mcvar}

Let $\bC_i=\rbra{Z_{i,t},\dots,Z_{i,t-k}}^\top$. To get $R_{\bZ_{t:(t-k)}}$ from $R_{Z_{i,t:(t-k)}}$ for $1\leq i\leq d$ and $R_{\bZ}$, one can firstly derive \[R_{\cbra{1,\dots,d}}=\corr\rbra{\rbra{Z_{1,t},\dots,Z_{d,t},\dots,Z_{1,t-k},\dots,Z_{d,t-k}}}
=\corr\rbra{\rbra{\bC_1^\top,\dots,\bC_d^\top}^\top}.\]
Then $R_{\bZ_{t:(t-k)}}$ can be obtained by reordering rows and columns of
$R_{\cbra{1,\dots,d}}$.
The definitions of $R_{\cbra{1,\dots,d}}$ and $R_{Z_{i,t:(t-k)}}$ for $1\leq i\leq d$ lead to
\begin{equation}\label{eq:correlation_matrix_of_Z1_to_Zn}
    R_{\cbra{1, \dots, d}} =
    \begin{pmatrix}
        R_{Z_{1,t:(t-k)}}       & \corr\rbra{\bC_1, \bC_2}       & \cdots & \corr\rbra{\bC_1, \bC_d} \\
        \corr\rbra{\bC_1, \bC_2}^T & R_{Z_{2,t:(t-k)}}           & \cdots & \corr\rbra{\bC_2, \bC_{d-1}} \\
        \vdots        & \vdots            & \ddots & \vdots \\
        \corr\rbra{\bC_1, \bC_d}^T & \corr\rbra{\bC_2, \bC_{d-1}}^T & \cdots & R_{Z_{d,t:(t-k)}}
    \end{pmatrix},
\end{equation}
where $\corr\rbra{\bC_i,\bC_j}$ for $1\leq i<j\leq d$ can be derived by the
formulas given below.

For $1\leq p \leq d$, let
\[
R_{Z_{p,t:(t-k)}} =
    \begin{pmatrix}
        1  & \varrho_{p,1}     & \cdots & \varrho_{p,k} \\
        \varrho_{p,1} & 1     & \cdots & \varrho_{p,k-1} \\
        \vdots           & \vdots             & \ddots & \vdots \\
        \varrho_{p,k} & \varrho_{p,k-1} & \cdots & 1
    \end{pmatrix}
\]
and
\begin{equation}
H_{p} =
 \begin{pmatrix}
  -I_{d_p} & \psi_{p,k} & \dots & \psi_{p,2} & \psi_{p,1} & 0 & 0 & \cdots & 0\\
  0 & -I_{d_p} & \psi_{p,k} & \dots & \psi_{p,2} & \psi_{p,1} & 0 & \cdots & 0 \\
  \vdots  & \ddots & \ddots & \ddots & \ddots & \ddots &   \ddots & \ddots & \vdots \\
  0   & \cdots & 0 & -I_{d_p} & \psi_{p,k} & \cdots & \psi_{p,2} & \psi_{p,1} & 0
 \end{pmatrix},
\end{equation}
where
\begin{equation}
\begin{pmatrix}
   \psi_{p,1} & \psi_{p,2} & \cdots & \psi_{p,k} \\
 \end{pmatrix}^\top =
 \begin{pmatrix}
  \varrho_{p,-k} & \varrho_{p,1-k}& \cdots & \varrho_{p,-1} \\
 \end{pmatrix}^\top
 \begin{pmatrix}
  1     & \varrho_{p,1}     & \cdots & \varrho_{p,k-1} \\
  \varrho_{p,1}   & 1     & \cdots & \varrho_{p,k-2} \\
  \vdots             & \vdots             & \ddots & \vdots \\
  \varrho_{p,k-1} & \varrho_{p,k-2} & \cdots & 1
 \end{pmatrix}^{-1},
\end{equation}
and $\varrho_{p,-m}=\varrho_{p,m}$.
Let $H_{p:(k+1)}$ denote the submatrix of the $(k+1)$-th block column of $H_p$ and $H_{p,:-(k+1)}$ denote the submatrix obtained by removing the $(k+1)$-th block column of $H_p$. If $A=(a_{ij})$ is an $m\times n$ matrix, the Kronecker product
$A\otimes B$ is defined as
\[
    \begin{pmatrix}
    a_{11}B & \cdots & a_{1n}B \\
    \vdots  & \ddots & \vdots \\
    a_{m1}B & \cdots & a_{mn}B \\
    \end{pmatrix}.
\]
Let $L$ be the
$(2k+1)$-dimensional exchange (or permutation) matrix whose elements in the anti-diagonal are one and all other elements are zero.
Let $\varrho_{ij,0}=\corr\rbra{Z_{i,t},Z_{j,t}}$ for $i\ne j$; this is
an element of $R_{\bZ}$. Then
\[
\corr\rbra{\bC_i, \bC_j} =
    \begin{pmatrix}
        \varrho_{ij,0}   & \varrho_{ij,1}   & \cdots & \varrho_{ij,k-1} \\
        \varrho_{ij,-1}  & \varrho_{ij,0}   & \cdots & \varrho_{ij,k-2} \\
        \vdots           & \vdots           & \ddots & \vdots \\
        \varrho_{ij,1-k} & \varrho_{ij,2-k} & \cdots & \varrho_{ij,0}
    \end{pmatrix},
\]
where
\begin{equation}
    \rbra{ \varrho_{ij, -k}, \dots, \varrho_{ij,-1}, \varrho_{ij,1}, \dots, \varrho_{ij,k} }^\top=
    -\varrho_{ij,0}\begin{pmatrix}
        H_{i,:-(k+1)}\\
        \rbra{H_jL}_{:-(k+1)}
    \end{pmatrix}^{-1}
    \begin{pmatrix}
        H_{i,:(k+1)}\\
        H_{j,:(k+1)}
    \end{pmatrix}.
\end{equation}

\section{Proof of \cref{pro:markov_order}}
\label{ap: proof}

\paragraph{Part 1 (The process is Markov of order $k$)}
Suppose the process changes from regime $g_b$ at time $t^*-1$
to regime $g'$ at time $t^*$.
Because $\{\bY_t\}$ is a Gaussian process,
$\bY_{t^*}$ is independent of $\bY_{t^*-\ell-1}$ given $\bY_{t^*-1}$ for any $\ell$ with $\ell\geq 1$, implying 
\begin{equation}\label{eq:con_dist1}
  \sbra{\bY_{t^*}\big|\bY_{t^*-1},\dots,\bY_{1}} \overset{\dee}{=} \sbra{\bY_{t^*}\big|\bY_{t^*-1}}.
\end{equation}
From the covariance matrix of conditional distributions of multivariate Gaussian random vectors,
\cref{eq:condition3} implies
\begin{equation} \label{eq:condition3.1}
  \begin{split}
   &\cov\rbra{\bY_{t^*+\tilde{\ell}}, \bY_{t^*-\ell-1}\big|\bY_{t^*-1}}
    = \cov\rbra{\bY_{t^*+\tilde{\ell}}, \bY_{t^*-\ell-1}} \\
   &\hspace{1cm}-\cov\rbra{\bY_{t^*+\tilde{\ell}},\bY_{t^*-1}}[\cov\rbra{\bY_{t^*-1},\bY_{t^*-1}}]^{-1}
    \cov\rbra{\bY_{t^*-1},\bY_{t^*-\ell-1}}
    =\bm{0},
  \end{split}
\end{equation}
for any $\ell, \tilde{\ell} \geq 1$.
Combining \cref{eq:condition2} and \cref{eq:condition3.1}, by induction,
\[
    \sbra{\bY_{t^*+\tilde{\ell}}\bot \bY_{t^*-\ell-1}}\big| \bY_{t^*+\tilde{\ell}-1},\dots,\bY_{t^*-1},
\]
i.e.,
\begin{equation}\label{eq:con_dist2}
  \sbra{\bY_{t^*+\tilde{\ell}}\big|\bY_{t^*+\tilde{\ell}-1},\dots,\bY_{1}} \overset{\dee}{=} \sbra{\bY_{t^*+\tilde{\ell}}\big|\bY_{t^*+\tilde{\ell}-1},\dots,\bY_{t^*-1}}.
\end{equation}
To prove that process $\{\bY_t\}$ is Markov of order $k$ need to show that the right hand side of \cref{eq:con_dist2} can be reduced to at most $k$ conditioning variables. Hence, only the situation when $\tilde{\ell} > k$ and $v_{t^*}=v_{t^*+1}=\cdots=v_{t^*+\tilde{\ell}}=g'$ needs to be considered.
Let
\[
    \bA_{t^*+\tilde{\ell}} &= \rbra{\bY^\top_{t^*+\tilde{\ell}-1}, \dots,
\bY^\top_{t^*+\tilde{\ell}-k+1}}^\top.
\]
Since the process is Markov of order $k-1=\max_{1\leq g\leq G}k_g$
within any regime,
\[
    \sbra{\bY_{t^*+\tilde{\ell}}\bot
\rbra{\bY^\top_{t^*+\tilde{\ell}-k},\dots,\bY^\top_{t^*}}^\top}\Bigm|\bA_{t^*+\tilde{\ell}}.
\]
Moreover, the condition in \cref{eq:condition3} implies $\bA_{t^*+\tilde{\ell}}\bot \bY_{t^*-1}$, and leads to
\[
    &\cov\rbra{\bY_{t^*+\tilde{\ell}}, \bY_{t^*-1}\big|\bA_{t^*+\tilde{\ell}}}\\
    =&\cov\rbra{\bY_{t^*+\tilde{\ell}}, \bY_{t^*-1}} -
    \cov\rbra{\bY_{t^*+\tilde{\ell}}, \bA_{t^*+\tilde{\ell}}}
  [\cov\rbra{\bA_{t^*+\tilde{\ell}}, \bA_{t^*+\tilde{\ell}}}]^{-1}
  \cov\rbra{\bA_{t^*+\tilde{\ell}}, \bY_{t^*-1}} \\
    =&\bm{0} - \cov\rbra{\bY_{t^*+\tilde{\ell}}, \bA_{t^*+\tilde{\ell}}}\cov^{-1}\rbra{\bA_{t^*+\tilde{\ell}}, \bA_{t^*+\tilde{\ell}}}\bm{0} = \bm{0}.
\]
According to \cref{eq:con_dist2}, it implies that if $\tilde{\ell}>k$, then
\begin{multline}\label{eq:con_dist3}
  \sbra{\bY_{t^*+\tilde{\ell}}\big|\bY_{t^*+\tilde{\ell}-1},\dots,\bY_{1}}
    \overset{\dee}{=}  \sbra{\bY_{t^*+\tilde{\ell}}\big|\bY_{t^*+\tilde{\ell}-1},\dots,\bY_{t^*-1}}
    \overset{\dee}{=}  \sbra{\bY_{t^*+\tilde{\ell}}\big|\bY_{t^*+\tilde{\ell}-1},\dots,\bY_{t^*+\tilde{\ell}-k+1}}.
\end{multline}
Note that \cref{eq:con_dist1}, \cref{eq:con_dist2}, and \cref{eq:con_dist3} hold regardless of the number of regime switches before $t^*$.

Since \cref{eq:condition1}, \cref{eq:condition2}, and \cref{eq:condition3} imply that the above holds for any regime switching time point,
$\cbra{\bY_t}_{t\in\ints}$ is Markov of order $k$.

\medskip

\paragraph{Part 2 (Closure under margins)}
Next, we show that \cref{eq:condition1}, \cref{eq:condition2}, and \cref{eq:condition3} imply that
the analogous conditions to these hold for any sub-process $\cbra{\bY_{I, t}}_{t\in\ints}$. {Let $P_I=\diag(\rho_i:i\in I)$.
Because $P$ is diagonal}, \cref{eq:condition1} implies
\begin{equation}\label{eq:uni_condition1}
    \cov\rbra{\bY_{I, t^*}, \bY_{I, t^*-1}} = P_I\cov\rbra{\bY_{I,t^*-1},\bY_{I,t^*-1}}.
\end{equation}
Combining \cref{eq:condition1} and \cref{eq:condition2} leads to
\begin{align} \label{eq:subset_cond}
    \cov\rbra{\bY_{t^*}, \bY_{t^*-\ell-1}}
    =& \cov\rbra{\bY_{t^*}, \bY_{t^*-\ell-1}\big|\bY_{t^*-1}} \\
    &+\cov\rbra{\bY_{t^*},\bY_{t^*-1}}
  [\cov\rbra{\bY_{t^*-1},\bY_{t^*-1}}]^{-1}
   \cov\rbra{\bY_{t^*-1},\bY_{t^*-\ell-1}}\\
    =&P R_{\bY,g}R^{-1}_{\bY,g}\cov\rbra{\bY_{t^*-1},\bY_{t^*-\ell-1}}
    \ =P\cov\rbra{\bY_{t^*-1},\bY_{t^*-\ell-1}}
\end{align}
for any $\ell\geq 1$.
Hence, $\cov\rbra{\bY_{I,t^*}, \bY_{I,t^*-\ell-1}}=
P_I\cov\rbra{\bY_{I,t^*-1},\bY_{I,t^*-\ell-1}}$
for $\ell\geq 1$.
Combining with \cref{eq:uni_condition1} leads to
\begin{align}
   &\cov\rbra{\bY_{I, t^*}, \bY_{I,t^*-\ell-1}\big| \bY_{I, t^*-1}} \\
    =& \cov\rbra{\bY_{I, t^*}, \bY_{I,t^*-\ell-1}}
  - \cov\rbra{\bY_{I, t^*}, \bY_{I,t^*-1}}
   [\cov\rbra{\bY_{I,t^*-1},\bY_{I,t^*-1}}]^{-1}
   \cov\rbra{\bY_{I,t^*-1}, \bY_{I,t^*-\ell-1}}  \\
    =& \cov\rbra{\bY_{I, t^*}, \bY_{I,t^*-\ell-1}} - P_I\cov\rbra{\bY_{I,t^*-1},\bY_{I,t^*-\ell-1}} = \bm{0},
\end{align}
i.e.,
\begin{equation}\label{eq:uni_condition2}
    \sbra{\bY_{I, t^*}\bot \bY_{I, t^*-\ell-1}}\big|\bY_{I, t^*-1},\quad \forall\ \ell\geq 1.
\end{equation}
Also, \cref{eq:condition3} implies:
\begin{equation}\label{eq:uni_condition3}
  \bY_{I, t^*+\tilde{\ell}} \bot \bY_{I, t^*-\ell},\quad \forall\ \ell, \tilde{\ell}\geq 1.
\end{equation}
Then, with the same procedure as in Part 1, combining \cref{eq:uni_condition1}, \cref{eq:uni_condition2}, and \cref{eq:uni_condition3} shows that $\cbra{\bY_{I, t}}_{t\in\ints}$ is Markov of order $k$. The model is closed under all margins.


\section{Baum-Welch algorithm}
\label{ap: baum-welch}

In this section, the Baum-Welch algorithm is given in the notation
of the model in \cref{def:model}.
All functions depend on $\btheta$ (the set of all parameters in the model)
as defined in \cref{sec:fitting_models}.

Let $\alpha_t$ be functions with $t$ inputs for $1\leq t\leq k$ and $k+1$ inputs for $t\geq k+1$, and they are defined as:
\begin{equation}
  \begin{split}
    \alpha_t(\omega_1,\dots,\omega_t;\btheta) &=f_{\bX_{t:1}, V_{t:1}}(\bx_t,\dots,\bx_1, \omega_t, \dots,\omega_1;\btheta)\quad\text{for } 1\leq t\leq k;
  \end{split}
\end{equation}
and
\begin{equation}
  \begin{split}
    \alpha_t(\omega_1,\dots,\omega_{k+1};\btheta) &=f_{\bX_{t:1},V_{t:(t-k)}}(\bx_t,\dots,\bx_1, \omega_{k+1}, \dots, \omega_1; \btheta)\quad\text{for } t\geq k+1.
  \end{split}
\end{equation}
It follows that
\[\alpha_1(\omega_1;\btheta) &= \Pr(V_1=\omega_1;\btheta)f_{\bX_1|V_1}(\bx_1|\omega_1;\btheta),\]
where $\Pr(V_1=\omega_1;\btheta)$ is the probability that the initial regime
is $\omega_1$,
$f_{\bX_1|V_1}$ is the conditional density of $\bX_1$ given $V_1$.
For $2\leq t\leq k$, 
\begin{equation}
  \begin{split}
    &\alpha_t(\omega_1,\dots,\omega_t;\btheta) \\
    =& f_{\bX_t|\bX_{(t-1):1}, V_{t:1}}(\bx_t|\bx_{t-1},\dots,\bx_1,\omega_t, \dots,\omega_1;\btheta)\\
    &\times f_{V_t|\bX_{(t-1):1}, V_{(t-1):1}}(\omega_t|\bx_{t-1},\dots,\bx_{1},\omega_{t-1}, \dots, \omega_{1};\btheta)\\
    &\times f_{\bX_{(t-1):1}, V_{(t-1):1}}(\bx_{t-1},\dots,\bx_{1},\omega_{t-1}, \dots,\omega_{1};\btheta)\\
    =& f_{\bX_t|\bX_{(t-1):1},V_{t:1}}(\bx_t|\bx_{t-1},\dots,\bx_1,\omega_t,\dots,\omega_1;\btheta)p(\omega_t|\omega_{t-1};\btheta)\alpha_{t-1}(\omega_1,\dots,\omega_{t-1};\btheta),
  \end{split}
\end{equation}
where $p(\omega_t|\omega_{t-1};\btheta)$ is the transition probability from
regime $\omega_{t-1}$ to regime $\omega_t$,
and for $t\geq k+1$,
\begin{equation}
  \begin{split}
    &\alpha_t(\omega_1,\dots,\omega_{k+1};\btheta) \\
    =& f_{\bX_t|\bX_{(t-1):1}, V_{t:(t-k)}}(\bx_t|\bx_{t-1},\dots,\bx_1,\omega_{k+1}, \dots,\omega_{1};\btheta)\\
    &\times f_{V_t|\bX_{(t-1):1}, V_{(t-1):(t-k)}}(\omega_{k+1}|\bx_{t-1},\dots,\bx_1, \omega_k, \dots, \omega_1;\btheta)\\
    &\times f_{\bX_{(t-1):1}, V_{(t-1):(t-k)}}(\bx_{t-1},\dots,\bx_1,\omega_k, \dots, \omega_{1};\btheta)\\
    =& f_{\bX_t|\bX_{(t-1):(t-k)}, V_{t:(t-k)}}(\bx_t|\bx_{t-1},\dots,\bx_{t-k},\omega_{k+1}, \dots,\omega_{1};\btheta)\\
    &\times \sum_g \sbra{ p(\omega_{k+1}|\omega_{k};\btheta)f_{\bX_{(t-1):1}, V_{(t-1):(t-k-1)}}(\bx_{t-1},\dots,\bx_1,\omega_k, \dots,\omega_{1}, g|\btheta)}\\
    =& f_{\bX_t|\bX_{(t-1):(t-k)}, V_{t:(t-k)}}(\bx_t|\bx_{t-1},\dots,\bx_{t-k},\omega_{k+1}, \dots,\omega_{1};\btheta)\times \sum_g p(\omega_{k+1}|\omega_{k};\btheta)\alpha_{t-1}(g,\omega_1,\dots,\omega_k;\btheta).
  \end{split}
\end{equation}

In the backward step, functions $\beta_t$ are defined in the similar manner:\\
for $1\leq t\leq k$,
\begin{equation}
    \beta_t(\omega_1,\dots,\omega_t;\btheta) = f_{\bX_{T:(t+1)}|\bX_{t:1},V_{t:1}}(\bx_{T},\dots,\bx_{t+1}| \bx_t,\dots,\bx_1, \omega_t, \dots, \omega_1;\btheta);
\end{equation}
for $t\geq k+1$,
\begin{equation}
  \beta_t(\omega_1,\dots,\omega_{k+1};\btheta)
  = f_{\bX_{T:(t+1)}|\bX_{t:(t-k)}, V_{t:(t-k)}}(\bx_T,\dots,\bx_{t+1}|\bx_t,\dots,\bx_{t-k}, \omega_{k+1}, \dots,\omega_1;\btheta).
\end{equation}
Then
\begin{equation}
  \beta_T(\omega_1,\dots,\omega_{k+1};\btheta) = 1,
\end{equation}
for $t \geq k+1$,
\begin{equation}
  \begin{split}
    &\beta_t(\omega_1,\dots,\omega_{k+1};\btheta) \\
    =& \sum_g f_{\bX_{T:(t+1)},V_{t+1}|\bX_{t:(t-k)}, V_{t:(t-k)}}(\bx_{T},\dots,\bx_{t+1},g|\bx_{t},\dots,\bx_{t-k}, \omega_{k+1}, \dots,\omega_{1};\btheta)\\
    =&\sum_g\biggl[
      f_{\bX_{T:(t+1)}|\bX_{t:(t-k)}, V_{(t+1):(t-k)}}(\bx_{T},\dots,\bx_{t+1}|\bx_{t},\dots,\bx_{t-k}, g, \omega_{k+1}, \dots,\omega_{1};\btheta)\\
    &\qquad\times
    f_{V_{t+1}|\bX_{t:(t-k)}, V_{t:(t-k)}}(r|\bx_{t},\dots,\bx_{t-k}, \omega_{k+1}, \dots,\omega_{1};\btheta)\biggr]\\
    =&\sum_r\biggl[
      f_{\bX_{T:(t+2)}|\bX_{(t+1):(t-k)}, V_{(t+1):(t-k)}}(\bx_{T},\dots,\bx_{t+2}|\bx_{t+1},\dots,\bx_{t-k}, g, \omega_{k+1}, \dots, \omega_{1};\btheta)\\
    &\qquad\times
    f_{\bX_{t+1}|\bX_{t:(t-k)}, V_{(t+1):(t-k)}}(\bx_{t+1}|\bx_{t},\dots,\bx_{t-k}, g, \omega_{k+1}, \dots,\omega_{1};\btheta)p(g|\omega_{k+1};\btheta)\biggl]\\
    =&\sum_g\biggl[
      \beta_{t+1}(\omega_2,\dots,\omega_{k+1},g;\btheta) \times p(g|\omega_{k+1};\btheta)\\
    &\qquad\times
    f_{\bX_{t+1}|\bX_{t:(t-k+1)}, V_{(t+1):(t-k+1)}}(\bx_{t+1}|\bx_{t},\dots,\bx_{t-k+1}, g, \omega_{k+1}, \dots, \omega_{2};\btheta) \biggl]
  \end{split}
\end{equation}
and for $t\leq k$,
\begin{equation}
  \begin{split}
    &\beta_t(\omega_1,\dots,\omega_{t};\btheta) \\
    =& \sum_g f_{\bX_{T:(t+1)},V_{t+1}|\bX_{t:1}, V_{t:1}}(\bx_{T},\dots,\bx_{t+1},g|\bx_{t},\dots,\bx_{1}, \omega_{t}, \dots,\omega_{1};\btheta)\\
    =&\sum_g\biggl[
      f_{\bX_{T:(t+1)}|\bX_{t:1}, V_{(t+1):1}}(\bx_{T},\dots,\bx_{t+1}|\bx_{t},\dots,\bx_{1}, g, \omega_{t}, \dots,\omega_{1};\btheta)\\
    &\qquad\times
    f_{V_{t+1}|\bX_{t:1}, V_{t:1}}(r|\bx_{t},\dots,\bx_1, \omega_{t}, \dots,\omega_{1};\btheta)\biggr]\\
    =&\sum_g\biggl[
      f_{\bX_{T:(t+2)}|\bX_{(t+1):1}, V_{(t+1):1}}(\bx_{T},\dots,\bx_{t+2}|\bx_{t+1},\dots,\bx_{1}, g, \omega_{t}, \dots,\omega_{1};\btheta)\\
    &\qquad\times
    f_{\bX_{t+1}|\bX_{t:1}, V_{(t+1):1}}(\bx_{t+1}|\bx_{t},\dots,\bx_{1}, g, \omega_{t}, \dots,\omega_{1};\btheta)p(g|\omega_{t};\btheta)\biggl]\\
    =&\sum_g\biggl[
      \beta_{t+1}(\omega_1,\dots,\omega_{t},g;\btheta) \times p(g|\omega_{t};\btheta)\times
    f_{\bX_{t+1}|\bX_{t:1}, V_{(t+1):1}}(\bx_{t+1}|\bx_{t},\dots,\bx_{1}, g, \omega_{t}, \dots,\omega_{1};\btheta) \biggl].
  \end{split}
\end{equation}
According to the definition, it follows that
\[\alpha_t(\omega_1,\dots,\omega_t;\btheta)\beta_t(\omega_1,\dots,\omega_t;\btheta)
  =f_{\bX_{1:T}, V_{1:t}}(\bx_1,\dots,\bx_T, \omega_1,\dots,\omega_t|\btheta)\]
for $1\leq t\leq k$, and
\[\alpha_t(\omega_1,\dots,\omega_{k+1};\btheta)\beta_t(\omega_1,\dots,\omega_{k+1};\btheta)
=f_{\bX_{1:T}, V_{(t-k):t}}(\bx_1,\dots,\bx_T,\omega_1,\dots,\omega_{k+1}|\btheta)\]
for $t\geq k+1$.
The likelihood of observations is
\begin{equation}
  f_{\bX_{1:T}}(\bx_1,\dots,\bx_T;\btheta) = \sum_{g=1}^G\alpha_1(g;\btheta)\beta_1(g;\btheta).
\end{equation}
It leads to that for $1\leq~t~\leq~k$:
\begin{equation}\label{gamma1}
  \begin{split}
     &\Pr(V_1=\omega_1,\dots,V_{t}=\omega_{t}|\bX_1=\bx_1,\dots,\bX_T=\bx_T;\btheta)\\
  = & \frac{\alpha_t(\omega_1,\dots,\omega_t;\btheta)\beta_t(\omega_1,\dots,\omega_t;\btheta)}
  {\sum_{g=1}^G\alpha_1(g;\btheta)\beta_1(g;\btheta)},
  \end{split}
\end{equation}
for $t\geq k+1$:
\begin{equation}\label{gamma2}
  \begin{split}
    & \Pr(V_{t-k}=\omega_1,\dots,V_{t}=\omega_{k+1}|\bX_1=\bx_1,\dots,\bX_T=\bx_T;\btheta)\\
  =&\frac{\alpha_t(\omega_1,\dots,\omega_{k+1};\btheta)\beta_t(\omega_1,\dots,\omega_{k+1};\btheta)}
  {\sum_{g=1}^G\alpha_1(g;\btheta)\beta_1(g;\btheta)}.
  \end{split}
\end{equation}

\end{appendices}

\bibliographystyle{apalike}

\begin{thebibliography}{}

  \bibitem[Biller, 2009]{biller2009copula}
  Biller, B. (2009).
  \newblock Copula-based multivariate input models for stochastic simulation.
  \newblock {\em Operations Research}, 57(4):\ 878--892.
  
  \bibitem[Biller and Nelson, 2003]{biller2003modeling}
  Biller, B. and Nelson, B. (2003).
  \newblock Modeling and generating multivariate time-series input processes
    using a vector autoregressive technique.
  \newblock {\em ACM Transactions on Modeling and Computer Simulation}, 13(3):\
    211--237.
  
  \bibitem[Chauvet and Papers, 2005]{chauvet2005}
  Chauvet, M. and Papers, N.~W. (2005).
  \newblock {\em Dating Business Cycle Turning Points}.
  \newblock National Bureau of Economic Research.
  
  \bibitem[Chauvet and Piger, 2008]{chauvet2008comparison}
  Chauvet, M. and Piger, J. (2008).
  \newblock A comparison of the real-time performance of business cycle dating
    methods.
  \newblock {\em Journal of Business \& Economic Statistics}, 26(1):42--49.
  
  \bibitem[Cheng, 2016]{Cheng2016}
  Cheng, J. (2016).
  \newblock A transitional {M}arkov switching autoregressive model.
  \newblock {\em Communications in Statistics. Theory and Methods},
    45(10):2785--2800.
  
  \bibitem[Hamilton, 1990]{Hamilton1990}
  Hamilton, J.~D. (1990).
  \newblock Analysis of time series subject to changes in regime.
  \newblock {\em Journal of Econometrics}, 45(1):39--70.
  
  \bibitem[Joe, 2014]{Joe2014}
  Joe, H. (2014).
  \newblock {\em Dependence Modeling with Copulas}.
  \newblock CRC Press, Boca Raton.
  
  \bibitem[Jones and Faddy, 2003]{Jones2003skew_t}
  Jones, M.~C. and Faddy, M.~J. (2003).
  \newblock A skew extension of the t-distribution, with applications.
  \newblock {\em Journal of the Royal Statistical Society. Series B, Statistical
    Methodology}, 65(1):159--174.
  
  \bibitem[McCracken and Ng, 2016]{Mccracken2016fred}
  McCracken, M.~W. and Ng, S. (2016).
  \newblock {FRED-MD}: A monthly database for macroeconomic research.
  \newblock {\em Journal of Business \& Economic Statistics}, 34(4):\ 574--589.
  
  \bibitem[Monbet and Ailliot, 2017]{Monbet2017}
  Monbet, V. and Ailliot, P. (2017).
  \newblock Sparse vector {M}arkov switching autoregressive models. application
    to multivariate time series of temperature.
  \newblock {\em Computational Statistics \& Data Analysis}, 108:\ 40--51.
  
  \bibitem[Sola and Driffill, 1994]{Sola1994}
  Sola, M. and Driffill, J. (1994).
  \newblock Testing the term structure of interest rates using a stationary
    vector autoregression with regime switching.
  \newblock {\em Journal of Economic Dynamics \& Control}, 18(3):601--628.
  
  \bibitem[Zhang et~al., 2023]{Zhang2023}
  Zhang, L., Joe, H., and Nolde, N. (2023).
  \newblock Margin-closed vector autoregressive time series models.
  \newblock {\em Journal of Time Series Analysis}.
  
  \bibitem[Zucchini et~al., 2017]{HMM2017}
  Zucchini, W., MacDonald, I.~L., and Langrock, R. (2017).
  \newblock {\em Hidden {M}arkov Models for Time Series: An Introduction Using
    {R}, Second Edition}.
  \newblock CRC Press.
\end{thebibliography}

\end{document}